\documentclass[12pt]{article}
\usepackage{amscd,amssymb,amsmath,latexsym,enumerate}
\usepackage[mathscr]{euscript}
\usepackage{epsfig}
\usepackage{fancybox}
\usepackage{verbatim}

\usepackage{color}
\newcommand{\red}{\textcolor[rgb]{0.50,0.00,0.00}} 
\title{Spectral flows associated to flux tubes}

\author{Giuseppe De Nittis, Hermann Schulz-Baldes
\\
\\
{\small Department Mathematik, Universit\"at Erlangen-N\"urnberg, Germany}
}

\date{ }

\newtheorem{theo}{Theorem}

\newtheorem{proposi}{Proposition}
\newtheorem{lemma}{Lemma}

\newcommand{\CM}{{\mathbb C}}
\newcommand{\NM}{{\mathbb N}}
\newcommand{\RM}{{\mathbb R}}
\newcommand{\SM}{{\mathbb S}}
\newcommand{\TM}{{\mathbb T}}
\newcommand{\ZM}{{\mathbb Z}}

\newcommand{\Aa}{{\cal A}}

\newcommand{\PP}{{\bf P}}

\newcommand{\EE}{{\bf E}}
\newcommand{\Bb}{{\cal B}}

\newcommand{\Tr}{\mbox{\rm Tr}}
\newcommand{\Tt}{{\cal T}}

\newcommand{\Cc}{{\cal C}}

\newcommand{\Kk}{{\cal K}}
\newcommand{\Hh}{{\cal H}}

\newcommand{\one}{{\bf 1}}

\newcommand{\SF}{{\rm Sf}} 
\newcommand{\Ind}{{\rm Ind}} 
\newcommand{\Ker}{{\rm Ker}}

\newcommand{\ph}{{\mbox{\rm\tiny ph}}}
\newcommand{\ess}{{\mbox{\rm\tiny ess}}}
\newcommand{\Maj}{{\mbox{\rm\tiny Maj}}}

\newcommand{\KPH}{K_{\mbox{\rm\tiny ph}}}
\newcommand{\KSL}{K_{\mbox{\rm\tiny sl}}}

\begin{document}

\maketitle

\begin{abstract}
When a flux quantum is pushed through a gapped two-dimensional tight-binding operator, there is an associated spectral flow through the gap which is shown to be equal to the index of a Fredholm operator encoding the topology of the Fermi projection. This is a natural mathematical formulation of Laughlin's Gedankenexperiment. It is used to provide yet another proof of the bulk-edge correspondence. Furthermore, when applied to systems with time reversal symmetry, the spectral flow has a characteristic $\ZM_2$ signature, while for particle-hole symmetric systems it leads to a criterion for the existence of zero energy modes attached to half-flux tubes. Combined with other results, this allows to explain all strong invariants of two-dimensional topological insulators in terms of a single Fredholm operator.
\end{abstract}


\section{Overview}
\label{sec-overview}

For the explanation of the quantum Hall effect, Laughlin suggested a Gedankenexperiment  during which an extra magnetic flux is inserted adiabatically into a two-dimensional system exposed to a constant magnetic field. This allows to argue for a quantized Hall conductance \cite{Lau}. Actually adiabatics is only needed to establish a connection to the Hall conductance and it is possible to understand the main topological insight of Laughlin's argument in purely spectral terms, namely as a spectral flow. For example, exactly $N$ states flow through the gap above the $N$th Landau level of the Landau operator as a flux is inserted, as can be seen by explicit calculation when modeling the singular flux either by the Aharanov-Bohm gauge or by adequate half-line boundary conditions \cite{AP}. Viewed from the perspective of \cite{ASS} also taken in the present paper, the spectral flow $N$ through the $N$th gap of the Landau operator is equal to the Chern number of the associated Fermi projection (on the lowest $N$ Landau bands) which in turn can be calculated as the index of the Fredholm operator
\begin{equation}
\label{eq-Fred}
T\;=\;P\,F\,P
\;,
\qquad
F\;=\;
\frac{X_1+\imath\,X_2}{|X_1+\imath\,X_2|}
\;,
\end{equation}
where $X_1$, $X_2$ are the components of the position operator and $P$ is the Fermi projection. Indeed, this operator is well-known to be Fredholm if the Fermi level lies in a gap (or even in a region of dynamical localization \cite{BES}), namely its kernel and cokernel are finite dimensional so that its index $\Ind(T)=\dim(\Ker(T))-\dim(\Ker(T^*))$ is well-defined. In this manner, the Laughlin argument appears as a special case of the general connection between the index of a given Fredholm operator and the spectral flow of a wide class of associated unitary dilations, as outlined in Appendix~A following Phillips work \cite{Phi} which is also rederived in a companion paper \cite{DS0}. Once this perspective is taken, the Laughlin argument acquires a remarkable stability and is not based on any explicit calculation as in \cite{AP}. Here it is presented for gapped tight-binding models with constant magnetic fields and with basically arbitrary hopping elements and potentials (Theorem~\ref{theo-LaughlinFlux}). While such a statement, even in the natural generality presented below, is a folk theorem both in the physics and mathematical physics communities \cite{ASS,BES}, a detailed proof does not seem to be in the literature. Closest (but not identical and actually slighly weaker) to ours is a statement in an unpublished manuscript of Macris \cite{Mac}, however, there the proof again involves adiabatics which in our opinion is unnatural due to the comments above. Here the general theorem from \cite{Phi,DS0} connecting index to spectral flow is applied, and as preparation a careful analysis of the magnetic translations associated to constant magnetic fields perturbed by a flux tube is carried out. These operators lie in a certain extension of the rotation algebra by the compact operators, see Appendix~B. It seems perceivable to us that there exists an extension of Theorem~\ref{theo-LaughlinFlux} to operators with no gap at the Fermi level, but for which the Fermi level lies in a region of dynamical Anderson localization. However, already the statement of such an extension would require a carefully formulated definition of spectral flow (presumably using finite volume approximations) and this goes beyond the scope of this work.

\vspace{.2cm}

As an application of Theorem~\ref{theo-LaughlinFlux}, a short and intuitive proof of the Elbau-Graf version \cite{EG} of the bulk-edge correspondence is given in Theorem~\ref{theo-bulkedge} in Section~\ref{sec-bulkedge}. The basic idea of the argument is due to Macris \cite{Mac}, but the details in that manuscript are flawed at several crucial points (in particular, the proof of his Lemma~2) and several simplifications are made here ({\it e.g.} the gauges for flux tubes are chosen differently). As this argument is based on spectral flow, it is presently not clear how to adapt it for a proof of the bulk-edge correspondence in presence of a mobility gap as stated in \cite{EGS}. It is then shown in Section~\ref{sec-covariant} how Theorem~\ref{theo-bulkedge} also implies one of the main results of \cite{SKR,KRS} which concerns the bulk-edge correspondence for families of covariant operators.  Needless to say, even though these arguments circumvent the use of $K$-theory and cyclic cohomology as explained in \cite{KRS}, we believe the $K$-theoretic interpretation to be of great conceptual interest and value. 

\begin{table}
\begin{center}
\begin{tabular}{|c|c|c||c||c|c|c|c|}
\hline
TRS&PHS &SLS & CAZ &  STI  & Effect
\\\hline\hline
$0$ &$0$&$0$& A    & $\ZM$  & QHE
\\
$0$&$0$&$ 1$ & AIII  &  & 
\\
\hline\hline
$0$ &$+1$&$0$ & D  & $\ZM$ & TQHE, ZM
\\
$-1$&$+1$&$1$  & DIII &  $\ZM_2$   & SCS, DZM
\\
$-1$&$0$&$0$ & AII &  $\ZM_2$  & QSHE, SCS
\\
$-1$&$-1$&$1$  & CII  &  & 
\\
$0$ &$-1$&$0$ & C & $2\,\ZM$  & SQHE
\\
$+1$&$-1$&$1$  &  CI &  & 
\\
$+1$&$0$&$0$ & AI  & &
\\
$+1$&$+1$&$1$  & BDI  &  & 
\\
\hline
\end{tabular}
\end{center}
\caption{{\it List of symmetry classes ordered by {\rm TRS}, {\rm PHS} and {\rm SLS} as well as the {\rm CAZ}-label. Then follow, for dimension $d=2$, the possible values of the {\rm STI}, and finally the physical effects tied to them. A more detailed discussion is given in the text.
\label{table1}
}}
\end{table}

\vspace{.2cm}

Section~\ref{sec-symmetries} discusses the fate of index of the Fredolm operator $T$ and of the Laughlin argument in systems which have supplementary discrete symmetries, namely time reversal symmetry (TRS), particle hole symmetry (PHS) and/or sublattice symmetry (SLS, also called a chiral symmetry). The TRS and PHS can be either even or odd and combinations of all three  symmetries lead to the so-called ten universality or symmetry classes which, following Altland and Zirnbauer \cite{AZ}, are often labelled by a corresponding Cartan label denoted by CAZ in Table~\ref{table1}. The theory of topological insulators \cite{SRFL,Kit} distinguishes different topological ground states within the CAZ classes. These topological phases are labelled by the so-called strong toplogical invariants (STI) which are usually understood by $K$-theory \cite{Kit,FM,Tan}. As will be discussed below, these $K$-theoretic invariants can take values either in $\ZM$ or $\ZM_2$. For the case of two-dimensional systems these STI are listed in Table~\ref{table1}. In this work, the $K$-theoretic point of view is not further developed, but rather a complementary concrete approach for the labelling of the phases is proposed. Actually, the remarkable fact is that all values of the STI can be computed by analyzing merely {\it one} Fredholm operator, namely $T$ defined in \eqref{eq-Fred}. This is possible because the various physical symmetries lead to symmetries of the Fredholm operator showing that its index is an arbitrary integer in Class A and D, an even integer in Class C, and vanishes in the other class, but has a $\ZM_2$ index as a secondary invariant in Class DIII and AII. To explain this in detail is the object of Section~\ref{sec-symmetries}. It can be summarized as follows. 

\vspace{.2cm}

\noindent {\bf Classification Scheme} {\it Suppose that the Fermi level lies in a region of dynamical localization in the sense of} \cite{BES}. {\it In each of the} CAZ {\it classes, the strong invariant of} \cite{SRFL,Kit} {\it can be calculated as the index $\Ind(T)$ or the $\ZM_2$-index $\Ind_2(T)=\dim(\Ker(T))\,\mbox{\rm mod}\,2$ of the Fredholm operator $T$ given in} \eqref{eq-Fred}. {\it If there is a gap at the Fermi level, all these indices can furthermore be calculated as a spectral flow in the spirit of the Laughlin argument. }

\vspace{.2cm}

\noindent Let us give some further explanation as to what the STI actually are in translation invariant and periodic systems, based on \cite{Kit,FM,Tan}. In Class A and AIII, the STI are given by the complex $K$-groups $K_0(C_0(\RM^2))=\ZM$ and $K_1(C_0(\RM^2))=0$ where $\RM^2$ is to be interpreted as the two-dimensional momentum space. As the one-point compactification of $\RM^2$ is the sphere $\SM^2$, the group $K_j(C_0(\RM^2))$ coincides with the reduced $K$-group $\widetilde{K}_j(C(\SM^2))$. In the tight-binding solid state systems analyzed in this work, the sphere $\SM^2$ should be replaced by the torus $\TM^2$ and this may (and does in some cases) produce supplementary so-called weak invariants \cite{Kit}, which are not analyzed here. These comments transpose verbatim to the remaining $8$ cases. There are Real $K$-groups $KR_j(C_0(\RM^2)_\tau)$ introduced in \cite{Ati} where $\tau$ is the involution induced by $\tau(k)=-k$ for $k\in\RM^2$, stemming from complex conjugation in physical space, and $j=0,\ldots, 7$. Again $KR_j(C_0(\RM^2)_\tau)\cong \widetilde{KR}_j(C_0(\SM^2)_\tau)$. These $KR$ groups are well-known to be $0,0,\ZM,\ZM_2,\ZM_2,0,2\,\ZM,0$ for $j=0,\ldots, 7$ respectively. By the above classification scheme, these values correspond again precisely to the possible values of the index of $T$. Let us stress though that the above classification scheme based on the invariants of $T$ ($\Ind$ and $\Ind_2$) applies to systems with broken translation invariance and merely requires dynamical localization which by \cite{BES} assures that $T$ is indeed a Fredholm operator. The groups $0,0,\ZM,\ZM_2,\ZM_2,0,2\,\ZM,0$ are also the homotopy groups (modulo Bott periodicity) of the classifying spaces for Real $K$-theory, given by skew-adjoint Fredholm operators on a real Hilbert space \cite{AS}. This connection will be further discussed in an up-coming work which will also contain an extension of the classification scheme to other dimensions. 

\vspace{.2cm}

Let us now discuss case by case the invariants of $T$ in some more detail, together with the associated physical effects. This list is also a summary of the main results of Section~\ref{sec-symmetries}.

\begin{itemize}

\item Class A contains systems without further symmetries and thus, in particular, electronic systems which exhibit a quantum Hall effect {\rm (QHE)}. This is already discussed above. The operator $T$ has no particular symmetry and $\Ind(T)$ can take arbitrary integer values.

\item Chiral unitary systems (Class AIII) have a vanishing spectral flow in dimension $d=2$.  Here $T$ has vanishing index, and no secondary invariant.

\item Class D contains Bogoliubov-de Gennes (BdG) Hamiltonians with even PHS, but no further symmetry. This symmetry does not imply any particular symmetry of the Fredholm operator $T$ though, and rather connects it to its conjugate Fredholm operator $(\one-P)F(\one-F)$. Hence the spectral flow and $\Ind(T)$ can take any integer value. For covariant operators, these integers are equal to the Chern number of the Fermi projection which in turn appear in the Kubo formula for the thermal quantum Hall effect {\rm (TQHE)} as a prefactor in the Wiedemann-Franz law \cite{VMFT}. Furthermore, in these systems an inserted half-flux quantum is of physical interest as it models a vortex of the pair creation field. The operator at half-flux has again an even PHS. Attached to these vortices are zero modes (ZM) whenever $\Ind(T)$ is odd. In second quantization the associated creation operators are self-adjoint so that one also speaks of Majorana modes.  While this fact is common knowledge in the physics community \cite{RG}, also for tight-binding models  \cite{Roy,EF}, Theorem~\ref{theo-PHS} seems to provide the first mathematical proof and also establishes the stability of these zero modes for a wide class of operators, containing {\it e.g.} random perturbations.

\item Class C contains BdG Hamiltonians with odd PHS and all the above statements of Class D hold. The physical effect in Class C systems is the spin quantum Hall effect (SQHE) \cite{RG}, and $\Ind(T)$ is actually equal to the spin Hall conductance as given by the Kubo formula \cite{RG}. The crucial difference w.r.t. Class D is that $\Ind(T)$ is always even in Class C systems (Theorem~\ref{theo-ClassC}).  Let us stress that this evenness is {\em not} related to the fact that Class C models appear as a pair when obtained as SU$(2)$-invariant models of a Class D system (as in \cite{AZ}). Our claim is that each of these two Class C models already has an even index, so the even nature is topological as also noted in \cite{SRFL,Kit}, see Table~\ref{table1}.  This has important implications for the zero modes. Actually, due to the evenness of $\Ind(T)$ such zero modes are {\em not} stable in Class C, other than claimed in \cite{Roy}.

\item Systems in Class AII have an odd TRS (half-integer spin). In this class, the most prominent toy model with non-trivial topology is the Kane-Mele model \cite{KM}, and it has a $\ZM_2$ invariant. The physical effects associated to it are the quantum spin Hall effect {\rm (QSHE)} and a spin-charge separation (SCS) \cite{QZ,RVL}. It was shown in \cite{Sch} that the odd TRS implies that the Fredholm operator $T$ is odd symmetric (in a sense recalled below) and therefore $\Ind_2(T)=\dim(\Ker(T))\;\mbox{\rm mod}\,2\in\ZM_2$ is a well-defined secondary invariant (the index $\Ind(T)$ itself vanishes). Indeed, it is shown in Section~\ref{sec-TRS} that $\Ind_2(T)=1$ for the Kane-Mele model. Theorem~\ref{theo-Z2TRS} then shows that such a non-trivial index leads to a characteristic spectral flow, which is intimately related to spin-charge separation \cite{QZ}. This theorem follows from a general result on the spectral flow of dilations of odd symmetric Fredholm operators, proved in \cite{DS0} and recalled in Appendix~\ref{sec-SFdilations}.

\item Class DIII comprises models with even PHS and odd TRS. These models inherit from Class AII the possibility to have non-trivial $\ZM_2$ indices. Indeed, it is shown in Section~\ref{sec-chiral} how models with such non-trivial topology can be constructed by a doubling procedure, similar as the Kane-Mele model is obtained from two Haldane models. Theorem~\ref{theo-DIII} states that non-trivial topology leads to Kramers' degenerate double zero modes (DZM) at half flux. In principle, also models in Class CII could have $\ZM_2$ invariants due to the odd TRS, but as the odd PHS already leads to even indices in Class C (Theorem~\ref{theo-ClassC}), this $\ZM_2$ index is trivial.

\item In the remaining Classes CI, AI and BDI the even TRS implies that the Fredholm operator $T$ is even symmetric in the sense of \cite{Sch} and thus $\Ind(T)=0$ and there is no naturally associated secondary invariant because all Fredholm operators with the corresponding symmetry lie in one connected component (see Theorem~5 in \cite{Sch}).

\end{itemize}


\section{Gauges for flux tubes}
\label{sec-fluxtubes}

The purpose of this section is to write out explicit formulas for two gauges of a flux tube though one cell of the square lattice $\ZM^2$. One is a discrete version of the standard Aharonov-Bohm gauge, the other one has the vector potential concentrated on a half-line and has already been used in other works \cite{Mac,LZX,EF}. These gauges have different properties which are crucial for the arguments below. As a preparation, some generalities about vector potentials, magnetic fields and gauges on the square lattice are collected in Section~\ref{sec-vectorpotentials}.

\subsection{Magnetic potentials, magnetic fields and gauge transformations}
\label{sec-vectorpotentials}

\begin{figure}
\begin{center}
\includegraphics[height=5.7cm]{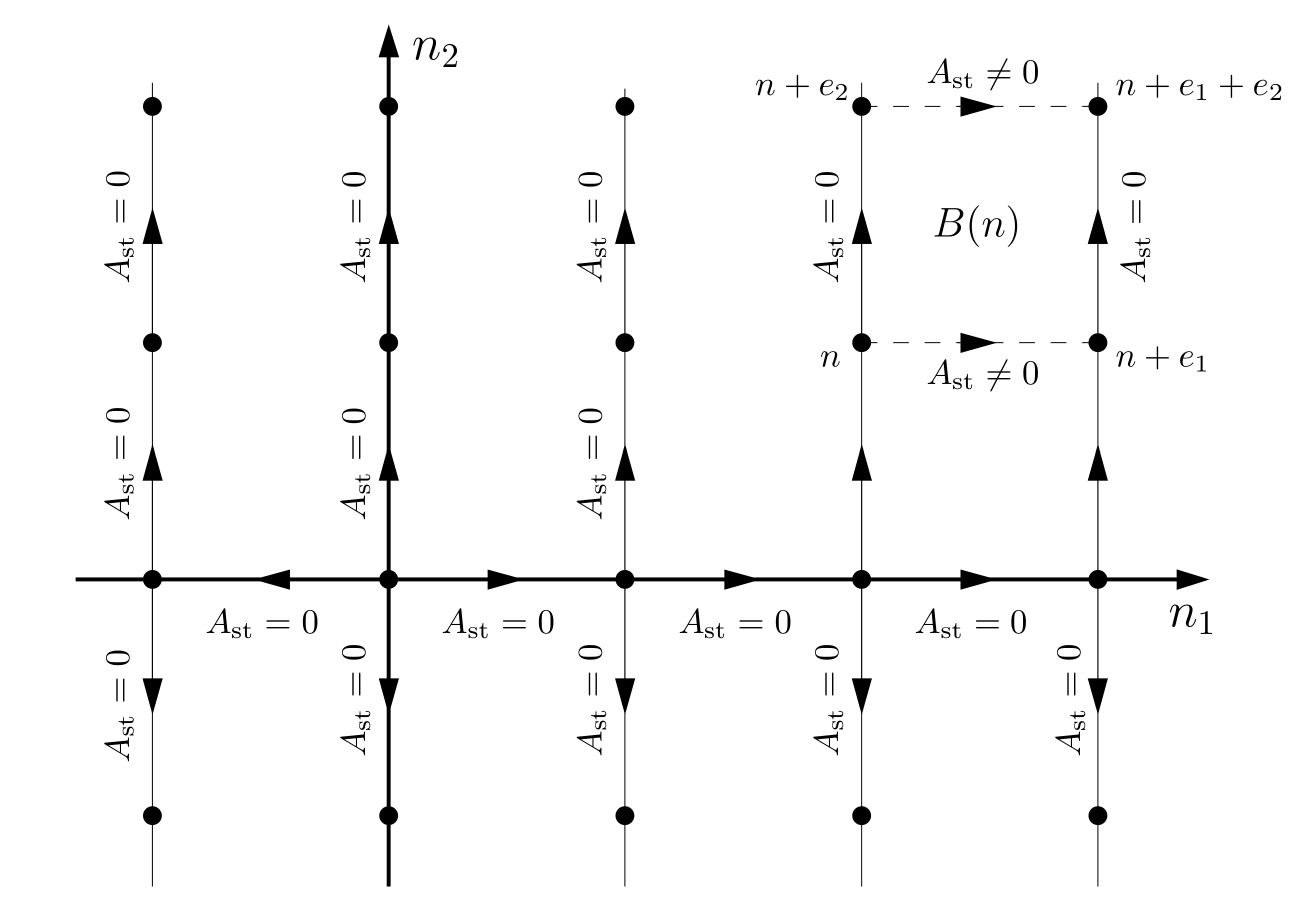}
\caption{\it 
Spanning tree on the lattice $\ZM^2$ for the standard gauge $A_{\mbox{\rm\tiny st}}$.
\label{fig-spantree}
}
\end{center}
\end{figure}

Let us view $\ZM^2$ as the vertices of an oriented graph, with oriented edges given by the  line segments $[n,n+e_j]$ between nearest vertices. Here $n\in\ZM^2$, and $e_1=(1,0)$ and $e_2=(0,1)$ denote the two unit vectors of $\ZM^2$. A magnetic potential on an oriented graph is a real-valued function on the oriented edges, hence in the present case a function $A:\ZM^2\times\ZM^2\to{\RM}$ satisfying $A(n,m)=0$ for $|n-m|\not = 1$ and $A(m,n)=-A(n,m)$. Associated to the magnetic potential $A$ is a magnetic field $B_A(n)\in{\RM}$ through the cell $(n,n+e_1,n+e_1+e_2,n+e_2)$ attached to the upper right at $n$, see Figure~\ref{fig-spantree}:
$$
B_A(n)
\;=\;
A(n,n+e_1)+A(n+e_1,n+e_1+e_2)+A(n+e_1+e_2,n+e_2)+A(n+e_2,n)
\;.
$$
This can be interpreted as the holonomy of $A$ along the path around the cell. Only $A$ and $B_A\,\mbox{\rm mod}\,2\pi$ will be relevant, but it will be convenient to maintain real values. Let us point out that the map $A\mapsto B_A$ is linear, namely $B_{A+\alpha A'}=B_A+\alpha B_{A'}$. If $x\in\RM^2\mapsto A(x)\in\RM^2$ is a (conventional) vector potential in continuous two-dimensional space, then an associated discretized magnetic potential on $\ZM^2$ in the above sense is obtained by the line integrals: 
\begin{equation}
\label{eq-gaugeint}
A(n,n+e_j)
\;=\;
\int^{n+e_j}_n dx\cdot A(x)
\;.
\end{equation}
Any magnetic field can be realized by a magnetic potential and two magnetic potentials realizing the same magnetic field are gauge transformation of each other, as shows the following result. 

\begin{proposi}
\label{prop-gauge}
{\rm (i)} Given $B:\ZM^2\to{\RM}$, there exists a magnetic potential $A$ such that $B=B_A$. 

\vspace{.1cm}

\noindent {\rm (ii)}  If $A$ and $A'$ are two magnetic potentials on $\ZM^2$ satisfying $B_A=B_{A'}$, then there exists a so-called gauge transformation $G:\ZM^2\to{\RM}$  such that
\begin{equation}
\label{eq-gaugetrafo}
A'(n,m)\;=\;A(n,m)+G(n)-G(m)
\;,
\qquad
|n-m|=1
\;.
\end{equation}
\end{proposi}

\noindent {\bf Proof.} It is known ({\it e.g.} \cite{CTT}) that a vector potential can be constructed using a spanning tree for the lattice. For sake of concreteness, let us choose one such tree (see Fig.~1) which then leads to what we call the standard gauge
$$
A_{\mbox{\rm\tiny st}}(n,n+e_j)
\;=\;
\delta_{j,1}\;\left(\delta_{n_2<0}\sum_{k=1}^{|n_2|}B(n_1,-k)\;-\; \delta_{n_2>0}\sum_{k=0}^{n_2-1}B(n_1,k)\right)
\;,
\qquad
n=(n_1,n_2)\in\ZM^2
\;.
$$
(ii) Choose $G(n)$ as the sum of $A'-A$ along a path from $0$ to $n$. As $B_{A'-A}=0$, this is independent of the choice of the path. 
\hfill $\Box$

\vspace{.2cm}

Next let us introduce the magnetic translations operators $S^A_1$ and $S^A_2$ on $\ell^2(\ZM^2)$  associated to the magnetic potential $A$:
\begin{equation}
\label{eq-magtransdef}
S^A_j\;=\; e^{\imath {A}(X+e_j,X)}S_j
\;=\;\sum_{n\in\ZM^2}
e^{-\imath A(n-e_j,n)}\,|n-e_j\rangle \langle n|
\;=\; S_je^{-\imath {A}(X-e_j,X)}
\;,
\qquad
j=1,2
\;.
\end{equation}
With $A$ given by \eqref{eq-gaugeint}, this is precisely the formula given in \cite[Theorem 3.2]{Ar2}. From the definition of $S^A_j$, it is clear that indeed only the values of $A\;\mbox{\rm mod}\,2\pi$ are relevant for the magnetic translations. On the other hand, $S^A_j$ depends on the choice of the magnetic potential $A$ and not only on $B$. The following commutation relation states that a phase factor given by the magnetic field is recovered by circulating around one cell:
\begin{equation}
\label{eq-magtransprop}
{S^A_2\,S^A_1\;(S^A_2)^*\,(S^A_1)^*
\;=\;
e^{\imath B_A(X)}}
\;,
\end{equation}
where $B_A(X)$ denotes the self-adjoint multiplication operator defined by $B_A(X)|n\rangle=B_A(n)|n\rangle$. The second main property of magnetic translations is their behavior under the gauge transformation given in \eqref{eq-gaugetrafo}:
\begin{equation}
\label{eq-magtransgauge}
S^{A'}_j
\;=\;
e^{-\imath \,G(X)}\,S^A_j\,e^{\imath \,G(X)}
\;,
\end{equation}
where $G(X)$ denotes again the multiplication operator given by $G$. Another property that is obvious from \eqref{eq-magtransdef} is its behavior under complex conjugation:
\begin{equation}
\label{eq-magtranscomplexconj}
\overline{S^A_j}
\;=\;
S^{(-A)}_j
\;.
\end{equation}

\subsection{Some explicit gauges}
\label{sec-explicitgauges}

Let us begin by recalling two standard gauges for a constant magnetic field $B(n)=B\in \RM$. The symmetric gauge ${A}_{\mbox{\rm\tiny sym}}$ and Landau gauge ${A}_{\mbox{\rm\tiny Lan}}$ are given by
%
$$
{A}_{\mbox{\rm\tiny Sym}}(n,n+e_j)
\;=\;
-\tfrac{1}{2}\,B\,n_2\,\delta_{j,1}\;+\;\tfrac{1}{2}\,B\,n_1\,\delta_{j,2}\;,
\qquad
{A}_{\mbox{\rm\tiny Lan}}(n,n+e_j)
\;=\;
-B\,n_2\,\delta_{j,1}\;.
$$
%
Note that ${A}_{\mbox{\rm\tiny Lan}}$ is actually the standard gauge used in Proposition~\ref{prop-gauge} and that the gauge transformation for the difference ${A}_{\mbox{\rm\tiny Sym}}-{A}_{\mbox{\rm\tiny Lan}}$ is given by $G(n)=-\tfrac{B}{2}n_1n_2$.

\vspace{.2cm}

Next let us consider the central object of this work, the discrete flux tube of flux $\alpha\in\RM$ through the cell $(m,m+e_1,m+e_1+e_2,m+e_2)$ attached to $m$. The magnetic field of this flux tube is ${B}(n)=2\pi\,\alpha\, \delta_{n,m}$. One possible gauge, termed {\it half-line} for sake of concreteness, is
$$
{A}_{\mbox{\rm\tiny HL}}(n,n+e_j)
\;=\;
-\,2\,\pi\,\alpha\;
\delta_{n_1,m_1}\,\delta_{n_2>m_2}\,
\delta_{j,1}
\;.
$$
A second gauge is obtained via \eqref{eq-gaugeint} from the standard singular Aharonov-Bohm gauge in $\RM^2$ attached at $m'=m+(\tfrac{1}{2},\tfrac{1}{2})$:
\begin{equation}
\label{eq-ABgauge0}
{A}_{\mbox{\rm\tiny AB}}(x)
\;=\;
\left(-\alpha\;\frac{x_2-m'_2}{(x_1-m'_1)^2+(x_2-m'_2)^2},\alpha\;\frac{x_1-m'_1}{(x_1-m'_1)^2+(x_2-m'_2)^2}\right)\;.
\end{equation}
Integration then leads to
\begin{equation}
\label{eq-ABgauge}
\begin{aligned}
{A}_{\mbox{\rm\tiny AB}}(n,n+e_j)
\; =\;&
-\;\alpha\left[\arctan\left(\frac{n_1+1-m'_1}{n_2-m'_2}\right)-\arctan\left(\frac{n_1-m'_1}{n_2-m'_2}\right)\right]\,\delta_{j,1}
\\
& +\;
\alpha\left[\arctan\left(\frac{n_2+1-m'_2}{n_1-m'_1}\right)-\arctan\left(\frac{n_2-m'_2}{n_1-m'_1}\right)\right]\,\delta_{j,2}
\;.
\end{aligned}
\end{equation}
Using the identity $\arctan(a)+\arctan(a^{-1})=\tfrac{\pi}{2}\,\tfrac{a}{|a|}$ for $a\in\RM/\{0\}$, it is indeed possible to check that the magnetic field associated with ${A}_{\mbox{\rm\tiny AB}}$ is exactly $2\pi\,\alpha\,\delta_{n,m}$. Alternatively, one can use the well-known properties of $A_{\mbox{\rm\tiny AB}}(x)$ to verify this. The gauge transformation in ${A}_{\mbox{\rm\tiny AB}}(n,m)-{A}_{\mbox{\rm\tiny HL}}(n,m)=\alpha(G(n)-G(m))$ is explicitly given by
%
\begin{equation}
\label{eq:gauge1}
G(n)\;=\;-\,\left[\pi\;\delta_{n_1>m'_1}\;+\;\arctan\left(\frac{n_2-m'_2}{n_1-m'_1}\right)\right]
\;.
\end{equation}

\subsection{Magnetic translations with flux tubes}
\label{sec-magtrans}

If $A={A}_{\mbox{\rm\tiny Sym}}$ with magnetic field $B$, the associated (Zak) magnetic translations are denoted by $S^B_j$. For $A=0$ so that $B=0$,  we also write $S_j$ instead of $S^B_j$. Next let us introduce the magnetic translations with constant magnetic field $B$ and flux tube $\alpha$ at $m$ by
\begin{equation}
\label{eq-magtransflux}
S^{B,\alpha}_j\;=\;S^A_j
\;,
\qquad
\mbox{where }\;\;A\,=\,{A}_{\mbox{\rm\tiny Sym}}\;+\;{A}_{\mbox{\rm\tiny AB}}
\;.
\end{equation}
In the following, note that there is a slight modification of the definition of $F$ w.r.t. \eqref{eq-Fred}.

\begin{proposi} 
\label{prop-magtransprop}
The operator differences $S^{B,\alpha}_j-S^B_j$ are compact. The operator functions $\alpha\in\RM\mapsto S^{B,\alpha}_j$ are norm continuous. Furthermore, 
\begin{equation}
\label{eq-Fphase}
S^{B,\alpha+1}_j
\;=\;
F\,S^{B,\alpha}_j\,F^*
\;,
\qquad
F\;=\;-\;\frac{X_1+m'_1+\imath (X_2+m'_2)}{|X_1+m'_1+\imath (X_2+m'_2)|}
\;,
\end{equation}
where $m'=(m'_1,m'_2)=m+(\frac{1}{2},\frac{1}{2})$. The commutators $[F,S^{B,\alpha}_j]$ are compact operators.
\end{proposi}

\noindent {\bf Proof.}  It follows from \eqref{eq-ABgauge} that $\lim_{|n|\to\infty}{A}_{\mbox{\rm\tiny AB}}(n+e_j,n)=0$. Thus $e^{\imath {A}_{\mbox{\rm\tiny AB}}{(X+e_j,X)}}-\one$ is a multiplication operator with factor decaying to $0$ at infinity and finitely degenerate eigenvalues, so that it is a compact operator. Due to \eqref{eq-magtransdef} this implies the first claim. As to the second it follows again from the relation \eqref{eq-magtransdef} and equation \eqref{eq-ABgauge} which shows that the gauge ${A}_{\mbox{\rm\tiny AB}}$ is linear in $\alpha$ with uniformly bounded coefficients, which is sufficient to insure the norm continuity. 

\vspace{.1cm}

To verify \eqref{eq-Fphase}, let us introduce the gauged magnetic shifts $Z^{B,\alpha}_j=e^{\imath \alpha G(X)}\, S^{B,\alpha}_j\,e^{-\imath \alpha G(X)}$ where $G$ is the gauge transformation given in \eqref{eq:gauge1}. Since the $\alpha$-dependance in  $Z^{B,\alpha}_j$ is given by the exponential in the {half-line} gauge ${A}_{\mbox{\rm\tiny HL}}$, one deduces that $Z^{B,\alpha+a}_j=Z^{B,\alpha}_j$ for all integers $a\in\ZM$. In particular, this implies that
$$
S^{B,\alpha+1}_j
\;=\;
e^{-\imath (\alpha+1)G(X)}\; Z^{B,\alpha}_j\;e^{\imath (\alpha+1)G(X)}
\;=\;
F\; S^{B,\alpha}_j\;F^*
$$
where the unitary $F$, written out using  \eqref{eq:gauge1}, is given by
$$
F
\;=\;
e^{-\imath G(X)}
\;=\;
-\,
\frac{X_1-m_1'}{|X_1-m_1'|}\; e^{\imath \arctan\left(\frac{X_2-m'_2}{X_1-m'_1}\right)}
\;=\;
-\,\frac{X_1+m'_1+\imath (X_2+m'_2)}{|X_1+m'_1+\imath (X_2+m'_2)|}
\;,
$$
where the following identities, holding for $x_1,x_2\in\RM/\{0\}$, were used
$$
\frac{x_1+\imath x_2}{|x_1+\imath x_2|}
\,=\,
\exp
\left(-
\imath\,\arctan\big(\frac{x_1}{x_2}\big)+\imath\,\frac{\pi}{2}\,\mbox{\rm sgn}(x_2)\right)
\,=\,
\exp
\left(
\imath\,\arctan\big(\frac{x_2}{x_1}\big)+\imath\,\frac{\pi}{2}\,(\mbox{\rm sgn}(x_1)-1)\right)
.
$$
For the last claim, let us rewrite using \eqref{eq-Fphase}
$$
[F,S^{B,\alpha}_j]
\;=\;
(S^{B,\alpha+1}_j-S^{B,\alpha}_j)F
\;=\;
(S^{B,\alpha+1}_j-S^{B}_j)F
\;+\;(S^{B}_j-S^{B,\alpha}_j)F
\;,
$$
so that the above allows to conclude.
\hfill $\Box$

\vspace{.2cm}

Appendix~\ref{sec-rotation} analyzes the C$^*$-algebra generated by $S^{B,\alpha}_1$ and $S^{B,\alpha}_2$. It is an extension of the rotation algebra by the compact operators and this allows to calculate its $K$-theory.

\vspace{.2cm}

\noindent {\bf Remark 1} The above proof shows how the unitary $F$ depends on the 
gauge transformation $G$ of  \eqref{eq:gauge1}. More generally, let us set
$$
F^\alpha\;=\;e^{-\imath \alpha G(X)}\;=\;e^{\imath\alpha\pi\delta_{X_1>m'_1}}
\; e^{\imath \alpha\arctan\left(\frac{X_2-m'_2}{X_1-m'_1}\right)}\;.
$$
With this notation, the relation between the translations $Z_j^{B,\alpha}$ in the  half-line gauge and the translations $S_j^{B,\alpha}$ in the Aharonov-Bohm gauge for $\alpha$ is given by $S_j^{B,\alpha}\;=\;F^{\alpha}\;Z_j^{B,\alpha}\;F^{-\alpha}$. From the definition $Z_j^{B,\alpha}=e^{\imath{A}_{\mbox{\rm\tiny HL}}}e^{\imath{A}_{\mbox{\rm\tiny Sym}}}S_j=e^{\imath{A}_{\mbox{\rm\tiny HL}}}S_j^{B,0}$ and the explicit form of ${A}_{\mbox{\rm\tiny HL}}$, it is evident that $Z_2^{B,\alpha}=S_2^{B,0}$ for all $\alpha\in\RM$ which, in particular, implies 
$S_2^{B,\alpha}=F^{\alpha}S_2^{B,0}F^{-\alpha}$. However, a similar relation is not true for $j=1$. In fact,
\begin{equation}
\label{eq:Rk01}
L_\alpha \;=\;Z_1^{B,\alpha}- S_1^{B,0}\;=\;\left(e^{\imath2\pi\alpha} -1\right)
\sum_{n_2>m_2}
\;e^{-\imath A_{\mbox{\rm\tiny Sym}}((m_1,n_2),(m_1+1,n_2))}\,|m_1,n_2\rangle \langle m_1+1,n_2|
\end{equation}
is non-vanishing and  not even compact, and one has $S_1^{B,\alpha}= F^{\alpha}\;(S_1^{B,0}+L_\alpha)\;F^{-\alpha}$. Hence inserting the flux $\alpha$ is {\em not} simply implemented by the unitary transformation  with $F^{\alpha}$, but it really introduces compact perturbations. More precisely, the algebra generated by $S_1^{B,\alpha}$ and $S_2^{B,\alpha}$ is a genuine extension of the rotation algebra (generated by the unperturbed magnetic translations $S_1^{B}$ and $S_2^{B}$) by the compact operators. This is explained in more detail in Appendix~\ref{sec-rotation}.
\hfill $\diamond$

\vspace{.2cm}

\noindent {\bf Remark 2} The claims of Proposition~\ref{prop-magtransprop} also hold if $A_{\mbox{\rm\tiny Sym}}$ is replaced by $A_{\mbox{\rm\tiny Lan}}$. On the other hand, replacing ${A}_{\mbox{\rm\tiny AB}}$ by ${A}_{\mbox{\rm\tiny HL}}$ is not allowed because the half-line gauge is actually a non-compact perturbation of the magnetic translations. Let us make this more explicit by analyzing the operator $\widetilde{S}^{B,\alpha}_j$ defined by
\begin{equation}
\label{eq-gaugechoice}
\widetilde{S}^{B,\alpha}_j\;=\;S^A_j
\;,
\qquad
\mbox{where }\;\;A\,=\,{A}_{\mbox{\rm\tiny Lan}}\;+\;{A}_{\mbox{\rm\tiny HL}}
\;.
\end{equation}
The difference between $Z_j^{B,\alpha}$ and $\widetilde{S}^{B,\alpha}_j$ is only in the choice for the gauge of the constant magnetic field $B$, while the gauge for the flux tube is concentrated on the half-line in both cases. Then $\widetilde{S}^{B,\alpha}_2=S_2$, but  $\widetilde{S}^{B,\alpha}_1$ is not compact perturbation of the magnetic translation in the Landau gauge 
given by $e^{-\imath BX_2}\,S_1$. Indeed, the replacement of $A_{\mbox{\rm\tiny Sym}}$  with $A_{\mbox{\rm\tiny Lan}}$ in \eqref{eq:Rk01} provides 
\begin{align*}
\widetilde{S}^{B,\alpha}_1
& = \; e^{-\imath BX_2}\,S_1\,S_1^*\,e^{\imath A_{\mbox{\rm\tiny HL}}(X+e_1,X)}\,S_1
\\
&=\;
e^{-\imath BX_2}\,S_1\;\left(\one\,+\,(e^{2\pi\imath\,\alpha}-1)\,\sum_{n_2>m_2}|m_1+1,n_2\rangle\langle m_1+1,n_2|
\right)
\;.
\end{align*}
In particular, $\widetilde{S}^{B,\alpha}_1-e^{-\imath BX_2}\,S_1$ is not compact. In spite of this unpleasant feature, the half-line gauge is of crucial importance in Section~\ref{sec-bulkedge}.
\hfill $\diamond$

\section{Spectral flow of the Laughlin argument}
\label{sec-Laughlin}

In this section, Hamiltonians on $\ell^2(\ZM^2)$ of the following form will be considered
\begin{equation}
\label{eq-disorderedHarper}
H_\alpha(\lambda)
\;=\;
\sum_{n=(n_1,n_2)\in\ZM^2}\,  t_{n}(\alpha)\;(S_1^{B,\alpha})^{n_1}\;(S_2^{B,\alpha})^{n_2}
\;+\;
\lambda\,V
\;,
\end{equation}
where $t_n(\alpha)\in\Kk^\sim$ (where $\Kk$ denotes the ideal of compact operators on $\ell^2(\ZM^2)$ and $\Aa^\sim$ is the unitalization of an algebra $\Aa$ obtained by adding multiples of the identity $\CM \one$) and $V=\sum_{n\in\ZM^2}v_n\,|n\rangle\langle n|$ is a uniformly bounded potential and  $\lambda\in[0,1]$ a coupling constant. It will be assumed that the hopping amplitudes decrease sufficiently fast so that
\begin{equation}
\label{eq-summability}
\sum_{n\in\ZM^2}\,\|t_n(\alpha)\|\;<\;
\infty
\;
\end{equation}
Moreover, for any $n\in\ZM^2$ the conditions
\begin{equation}
\label{eq-hoppingcond}
t_{-n}(\alpha)\;=\;(S_2^{B,\alpha})^{-n_2}\; (S_1^{B,\alpha})^{-n_1}\;t_n(\alpha)^*
\;(S_2^{B,\alpha})^{n_2}\;(S_1^{B,\alpha})^{n_1}\;,
\qquad
F t_n(\alpha) F^*\;=\;t_n(\alpha+1)
\;,
\end{equation}
are supposed to hold. They guarantee respectively $H_\alpha(\lambda)^*=H_\alpha(\lambda)$ and $FH_\alpha(\lambda)F^*=H_{\alpha+1}(\lambda)$.

\vspace{.2cm}

\noindent {\bf Remark} 
Definition~\eqref{eq-disorderedHarper}  combined with conditions \eqref{eq-hoppingcond} may seem a little unnatural at first glance.  However, Proposition~\ref{prop-magtransprop} implies that the commutator $[S_1^{B,\alpha},S_2^{B,\alpha}]$ is a non-vanishing compact operator when $\alpha\neq 0$, and thus the ordering of the magnetic shifts $S_1^{B,\alpha}$ and $S_2^{B,\alpha}$ becomes relevant. In \eqref{eq-disorderedHarper} a particular choice of ordering has been made and this requires \eqref{eq-hoppingcond}. Of course, if there are only nearest neighbor hopping terms like in the Harper Hamiltonian this is not an issue. Furthermore, for $\alpha=0$ the commutator is just a number which can be absorbed in $t_n(0)\in\CM\one$.
\hfill $\diamond$

\vspace{.2cm}

At $\lambda=1$, the Hamiltonian is simply denoted by $H_\alpha=H_\alpha(1)$, and for $\alpha=0$ the notations $H=H_0$ and $t_n=t_n(0)$ are used. If  $t_{e_1}=t_{-e_1}=t_{e_2}=t_{-e_2}=t\,\one$ with $t\in\RM$ are the only non-vanishing hopping amplitudes, then the Hamiltonian $H=H_0$ is the two-dimensional representation of the Harper Hamiltonian with constant magnetic flux $B$ through each unit cell, and $V$ allows, {\it e.g.}, to add a random potential or a compactly supported scattering-type potential.  Furthermore, in $H_\alpha$ the magnetic translations $S^{B,\alpha}_j$  given by \eqref{eq-magtransflux} add an extra flux through the unit cell attached at $m\in\ZM^2$.

\vspace{.2cm}

Let us begin by collecting a few basic mathematical properties of the Hamiltonian \eqref{eq-disorderedHarper} which follow rather directly from the properties of the magnetic translations.

\begin{proposi} 
\label{prop-fluxprepare} Let $g:\RM\to\CM$ be continuous. Then the following properties hold:

\vspace{.1cm}

\noindent {\rm (i)} $H_\alpha-H_0$ and $g(H_\alpha)-g(H_0)$ are compact.

\vspace{.1cm}

\noindent {\rm (ii)} $\sigma_\ess(H_\alpha)=\sigma_\ess(H_0)$

\vspace{.1cm}

\noindent {\rm (iii)} $g(H_{\alpha+1})=Fg(H_{\alpha})F^*$ with unitary $F$ as in {\rm \eqref{eq-Fphase}}.

\vspace{.1cm}

\noindent {\rm (iv)} $\sigma(H_{\alpha+1})=\sigma(H_\alpha)$

\vspace{.1cm}

\noindent {\rm (v)} The commutators $[F,g(H_\alpha)]$ are compact.

\vspace{.1cm}

\noindent {\rm (vi)} $\alpha\in\RM\mapsto g(H_\alpha)$ is norm continuous.

\vspace{.2cm}

\noindent All claims also hold for $H_\alpha(\lambda)$ with $\lambda\not = 1$.
\end{proposi}

\noindent {\bf Proof.} By Proposition~\ref{prop-magtransprop}, $S^{B,\alpha}_j-S^B_j$ is compact for $j=1,2$. Thus $(S_1^{B,\alpha})^{n_1}(S_2^{B,\alpha})^{n_2}-(S^B)^n$ is compact for all $n=(n_1,n_2)\in\ZM^2$ and the summability hypothesis \eqref{eq-summability} then implies that $H_\alpha-H_0$ is compact. Furthermore, telescoping
$$
(H_\alpha)^k-(H_0)^k
\;=\;
\sum_{l=1}^{k-1}\,(H_\alpha)^{k-l}(H_\alpha-H_0)(H_0)^{l}
$$
shows that also $(H_\alpha)^k-(H_0)^k$ is also compact for any $k\geq 1$ and combined with Weierstra{\ss} theorem and the norm closedness of the compact operators this implies (i).  By Weyl's theorem also (ii) follows. Furthermore \eqref{eq-Fphase} and $FVF^*=V$ lead first to $H_{\alpha+1}=FH_{\alpha}F^*$, and combined with Weierstra{\ss} approximation to (iii). Item (iv) is then a direct consequence of (iii), and (v) combines (iii) and (i): 
$$
[F,g(H_\alpha)]
\;=\;
\bigl(
g(H_{\alpha+1})-g(H_0)+g(H_0)-g(H_\alpha)
\bigr)F
\;\in\;\Kk
\;.
$$
Finally the continuity (vi) follows from the norm continuity of $S^{B,\alpha}$ stated in Proposition~\ref{prop-magtransprop}.
\hfill $\Box$

\vspace{.2cm}

The focus will now be on operators satisfying the following

\vspace{.2cm}

\noindent {\bf Gap hypothesis:} {\it The Fermi level $\mu\in\RM$ lies in a spectral gap of $H_0$ and in a gap of the essential spectrum of $H_0(\lambda)$ for all $\lambda\in[0,1]$.}

\vspace{.2cm}

\begin{figure}
\begin{center}
\includegraphics[height=6cm]{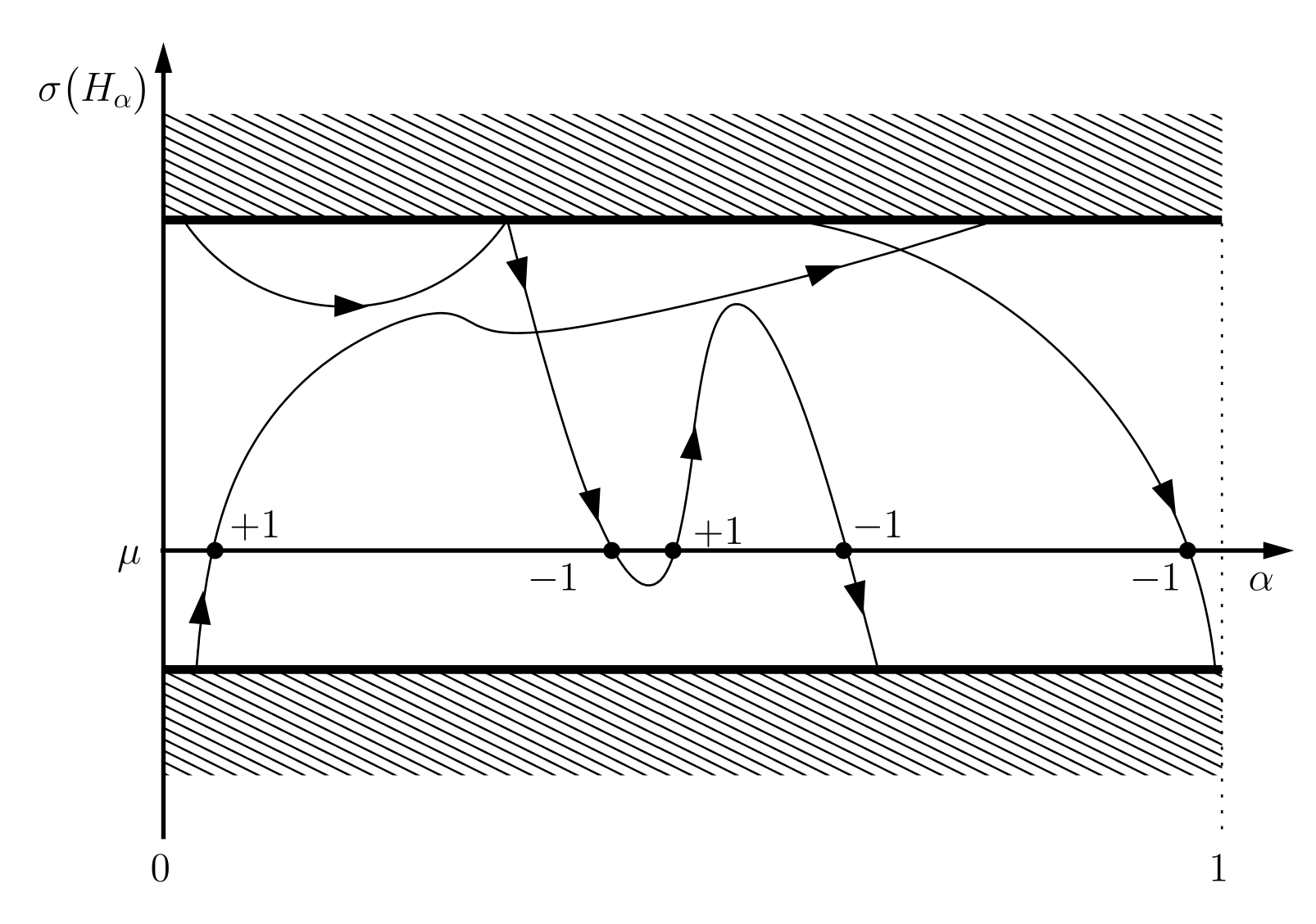}
\caption{\it 
Typical pattern of the spectral flow associated with a flux tube insertion $\alpha\mapsto H_\alpha$.
\label{fig-SF}
}
\end{center}
\end{figure}

\noindent The second part of the hypothesis can be slightly weakened by allowing also $\mu$ to depend on $\lambda$, but for sake of simplicity this is not written out here. Let us point out that the Gap hypothesis does not exclude that $H_0(\lambda)$ has bound states close to $\mu$ (namely, discrete spectrum resulting {\it e.g.} from a compact potential $V$). Due to Proposition~\ref{prop-fluxprepare}, as function of $\alpha\in[0,1]$ only the discrete spectrum of $H_\alpha(\lambda)$ is changing and may lead to eigenvalues passing by $\mu$. In fact, these eigenvalues vary real analytically in $\alpha$ due to the analytic dependence of $H_\alpha(\lambda)$ on $\alpha$. The operators at $\alpha=0$ and $\alpha=1$  are isospectral by Proposition~\ref{prop-fluxprepare}. Counting the eigenvalue passages along the path weighted by the multiplicities and a positive or negative sign pending on whether the eigenvalues increase or decrease allows to define the integer valued spectral flow $\SF(\alpha\in[0,1]\mapsto H_\alpha\;\mbox{\rm by}\;\mu)$ by $\mu$. This is illustrate in Figure~\ref{fig-SF}. As here the eigenvalue curves of the discrete spectrum are real analytic, the intuitive notion of spectral flow indeed leads to mathematically sound definition. Let us note that one may suspect there to be a problem defining the spectral flow in case $\mu$ happens to be an eigenvalue of $H_0(\lambda)$, but actually there is no issue because $\alpha\in[0,1]\mapsto\sigma(H_\alpha(\lambda))$ is really a closed loop by Proposition~\ref{prop-fluxprepare}(iv) so that the flow by $\mu$ is well-defined. A definition of spectral flow for the more general case of norm continuous families of self-adjoint operators can be found in \cite{Phi,DS0}. These references also discuss further properties of the spectral flow, such as its homotopy invariance. In order to familiarize the reader with the notion of spectral flow, let us provide an alternative formula which will also be used below.

\begin{proposi} 
\label{prop-LaughlinFlux} 
Suppose that the closed interval $\Delta\subset\RM$ lies in a gap of $H$ and let $g:\RM\to[0,1]$ be a smooth non-increasing function which is equal to $1$ on the left of $\Delta$ and $0$ on the right of $\Delta$. Then for $\mu\in\Delta$
\begin{equation}
\label{eq-LaughlinFlow2}
\SF\bigl(\alpha\in[0,1]\mapsto H_\alpha\;\mbox{\rm by}\;\mu\bigr)
\;=\;
-\;\int_0^1 d\alpha\;
\Tr\bigl(
g'(H_\alpha)\,\partial_\alpha H_\alpha
\bigr)
\;.
\end{equation}
Also the r.h.s. is manifestly gauge invariant.
\end{proposi}

\noindent {\bf Proof.} Let $E_l(\alpha)$, $l=1,\ldots,L$, denote the finite number of eigenvalues of $H_\alpha$ lying in $\Delta$ with normalized eigenvectors $\psi_l(\alpha)$ where $L$ is the maximal number of eigenvalues in $\Delta$ for all $\alpha$'s. These eigenvalues and eigenvectors are real analytic in $\alpha$. As the support of $g'$ lies in $\Delta$, the operator 
$$
g'(H_\alpha)\;=\;\sum_{l=1}^L g'(E_l(\alpha))\,|\psi_l(\alpha)\rangle\langle\psi_l(\alpha)|
$$
is finite rank. By the Fundamental Theorem,
$$
\SF\bigl(\alpha\in[0,1]\mapsto H_\alpha\;\mbox{\rm by}\;\mu\bigr)
\;=\;
-\;\int^1_0 d\alpha\;\sum_{l=1}^L\;g'(E_l(\alpha))\;\partial_\alpha\,E_l(\alpha)
\;.
$$
Using $\partial_\alpha\,E_l(\alpha)=\partial_\alpha\,\langle\psi_l(\alpha)|H_\alpha|\psi_l(\alpha)\rangle$ one readily concludes the proof. For the final claim, let $\widetilde{H}_\alpha=e^{\imath G_\alpha(X)}H_\alpha e^{-\imath G_\alpha(X)}$ be the Hamiltonian in another gauge. Then
$$
\Tr\bigl(
g'(H_\alpha)\,\partial_\alpha H_\alpha
\bigr)
\;=\;
\Tr\bigl(
e^{-\imath G_\alpha(X)}g'(\widetilde{H}_\alpha)e^{\imath G_\alpha(X)}\,\partial_\alpha \bigl(e^{-\imath G_\alpha(X)}\widetilde{H}_\alpha e^{\imath G_\alpha(X)}\bigr)
\bigr)
\;=\;
\Tr\bigl(
g'(\widetilde{H}_\alpha)\,\partial_\alpha \widetilde{H}_\alpha
\bigr)
\;,
$$
where in the last equality the cyclicity of the trace is used to cancel out $2$ terms.
\hfill $\Box$

\vspace{.2cm}

The following theorem connecting the spectral flow to an index is the central result of this paper. Due to the preparations in Proposition~\ref{prop-fluxprepare}, it is a corollary of a general statement of \cite{Phi}, also proved in \cite{DS0}, and recalled in Appendix~\ref{sec-SFdilations}. 

\begin{theo} 
\label{theo-LaughlinFlux} 
Suppose the {\rm Gap hypothesis} holds and let $P_\mu(\lambda)=\chi(H_0(\lambda)\leq \mu)$ be the Fermi projection of $H_0(\lambda)$ on energies below $\mu$. Then $P_\mu(\lambda) FP_\mu(\lambda)$ is a Fredholm operator on $P_\mu(\lambda)\,\ell^2(\ZM^2)$ and for all $\lambda\in[0,1]$ its index is given by
\begin{equation}
\label{eq-LaughlinFlow}
\SF\bigl(\alpha\in[0,1]\mapsto H_\alpha(\lambda)\;\mbox{\rm by}\;\mu\bigr)
\;=\;
\Ind(P_\mu(\lambda) FP_\mu(\lambda))
\;.
\end{equation}
Moreover, these expression is constant in $\lambda\in[0,1]$.
\end{theo}

\noindent {\bf Proof.} 
By the Gap hypothesis there exists a continuous and non-increasing function $g:\RM\to [0,1]$ such that $P_\mu(\lambda)=g(H_0(\lambda))$.  Therefore by Proposition~\ref{prop-fluxprepare}(v) the operators $P_\alpha=g(H_\alpha(\lambda))$ have compact commutators  $[F,P_\alpha]$. This implies the claimed Fredholm property and the constancy of the index on the r.h.s. of \eqref{eq-LaughlinFlow} follows from the homotopy invariance of the index. Furthermore all the hypothesis of Theorem~\ref{theo-FredFlow} in Appendix~\ref{sec-SFdilations} are verified. Thus
\begin{equation}
\label{eq-SFintermed}
\SF\bigl(\alpha\in[0,1]\mapsto g(H_\alpha(\lambda))\;\mbox{\rm by}\;\tfrac{1}{2}\bigr)
\;=\;
\Ind(P_\mu(\lambda) FP_\mu(\lambda))
\;.
\end{equation}
However, the spectral flow on the l.h.s. is precisely equal to the spectral flow in \eqref{eq-LaughlinFlow}. 
\hfill $\Box$

\vspace{.2cm}

The following complement to Theorem~\ref{theo-LaughlinFlux}, used in the Section~\ref{sec-bulkedge} below, shows that one also may cut out finite portions of the physical space $\ZM^2$ without changing the spectral flow. Roughly reformulated, this means that also compactly supported, infinite potentials do not change the spectral flow.

\begin{proposi} 
\label{prop-LaughlinFluxFiniteVol} 
Suppose the {\rm Gap hypothesis} holds. For $\Lambda\subset\ZM^2$ set $\Pi_{\Lambda}=\sum_{n\in\Lambda}|n\rangle\langle n|$ and $\Pi_{\Lambda^c}=\one-\Pi_\Lambda$. Then, for $\Lambda$ finite and $H_\alpha^{\Lambda}=\Pi_{\Lambda^c}H_\alpha\Pi_{\Lambda^c}$, one has
\begin{equation}
\label{eq-LaughlinFlowFiniteVol}
\Ind(P_\mu FP_\mu)
\;=\;
\SF\bigl(\alpha\in[0,1]\mapsto H_\alpha^\Lambda\;\mbox{\rm by}\;\mu\bigr)
\;=\;
-\;\int_0^1 d\alpha\;
\Tr\bigl(
g'(H_\alpha^\Lambda)\,\partial_\alpha H_\alpha^\Lambda
\bigr)
\;,
\end{equation}
where $P_\mu=\chi(H\leq \mu)$ is the Fermi projection of $H$. The r.h.s. is still gauge invariant.
\end{proposi}

\noindent {\bf Proof.} First of all, $H_\alpha^{\Lambda}$ is again a compact perturbation of $H_\alpha$ so that the essential spectra coincide. Furthermore, the projection $\Pi_\Lambda$ commutes with $F$. Now the proof of the first equality is a modification of the proof of Theorem~\ref{theo-LaughlinFlux} using the homotopy $\beta\in[0,1]\mapsto (\one-\beta\Pi_\Lambda)H_\alpha(\one-\beta\Pi_\Lambda)$. The second equality follows by the same argument as Proposition~\ref{prop-LaughlinFlux}.
\hfill $\Box$

\vspace{.2cm}

Let us conclude this section by analyzing what happens if several flux tubes are inserted simultaneously. 

\begin{proposi} 
\label{prop-severalfluxes} 
Suppose the {\rm Gap hypothesis} holds. Let $\alpha\in[0,1]\mapsto H_{(\alpha)}$ denote the family of Hamiltonians with a flux $\alpha$ through the lattice cells attached to $m^{(1)},\ldots,m^{(L)}\in\ZM^2$. Then
$$
\SF\bigl(\alpha\in[0,1]\mapsto H_{(\alpha)}(\lambda)\;\mbox{\rm by}\;\mu\bigr)
\;=\;
L\;\Ind(P_\mu F P_\mu)
\;,
$$
where $P_\mu=\chi(H\leq \mu)$ is the Fermi projection on energies below $\mu$ lying in the gap.
\end{proposi}

\noindent {\bf Proof.} Associated to each $l=1,\ldots,L$ there is a Dirac phase $F_{(l)}$ defined as in formula \eqref{eq-Fphase}. Setting  $F'=F_{(1)}\cdots F_{(L)}$, one then verifies all the claims of Proposition~\ref{prop-fluxprepare}, in particular, the identity $H_{(1)}=F'H_{(0)}(F')^*$ as well as compactness of $H_{(\alpha)}-H_{(0)}$. Now the proof of Theorem~\ref{theo-LaughlinFlux} shows
$$
\SF\bigl(\alpha\in[0,1]\mapsto H_{(\alpha)}\;\mbox{\rm by}\;\mu\bigr)
\;=\;
\Ind(P_\mu F'P_\mu)
\;.
$$
Furthermore $[F_{(l)},P_\mu]$ is compact so that
$$
P_\mu F'P_\mu
\;=\;
(P_\mu F_{(1)}P_\mu)\cdots (P_\mu F_{(L)}P_\mu)
\;+\;K
\;,
$$
for some compact operator $K$. The multiplicative property of the index and its invariance under compact perturbations implies
$$
\Ind(P_\mu F'P_\mu)
\;=\;
\sum_{l=1}^L\;\Ind(P_\mu F_{(l)}P_\mu)
\;.
$$
But all of the indices on the r.h.s. are equal to $\Ind(P_\mu FP_\mu)$, concluding the proof.
\hfill $\Box$

\section{Flux tube proof of the bulk-edge correspondence}
\label{sec-bulkedge}

This section is about the half-space operator $\widehat{H}$ acting on $\ell^2(\ZM\times\NM)$ simply obtained by restriction from the Hamiltonian $H$ given in \eqref{eq-disorderedHarper} with $\alpha=0$ and $\lambda=1$. This corresponds to imposing Dirichlet boundary conditions on the half-plane $\ZM\times\NM$. In principle all the below also holds for other local boundary conditions, but this is not analyzed in detail (non-local boundary conditions like the spectral boundary conditions of Atiyah-Patodi-Singer are not allowed though). For sake of simplicity, it will be assumed that the sum in \eqref{eq-disorderedHarper} is finite by imposing the constraint $|n|\leq R$ for some finite range $R\in\NM$. The half-plane operator $\widehat{H}$ describes the boundary of a quantum Hall system and therefore there are chiral edge currents. The edge current density operator is defined as the commutator $\imath\, [\Pi_<,\widehat{H}]$ where the quarter plane projection $\Pi_<$ is given by
\begin{equation}
\label{eq-quarterplane}
\Pi_<
\;=\;
\sum_{n_1<0}\;\sum_{n_2>0}\;|n_1,n_2\rangle\langle n_1,n_2|
\;.
\end{equation}
The following theorem states that the boundary current density is well-defined and quantized by a number which depends on the topology contained in the Fermi projection of the Hamiltonian $H$ acting on the plane. 


\begin{theo} 
\label{theo-bulkedge} 
Suppose that the closed interval $\Delta\subset\RM$ lies in a gap of $H$ and let $g:\RM\to[0,1]$ be any smooth non-increasing function which is equal to $1$ on the left of $\Delta$ and $0$ on the right of $\Delta$. Then
\begin{equation}
\label{eq-bulkedgepointwise}
\Tr\bigl(g'(\widehat{H})\,\imath \,[\Pi_<,\widehat{H}]\bigr)
\;=\;
-\;\frac{1}{2\pi}\;\Ind(P_\mu FP_\mu)
\;,
\end{equation}
where as above $P_\mu=\chi(H\leq \mu)$ is the Fermi projection on energies below $\mu\in\Delta$.
\end{theo}

A version of Theorem~\ref{theo-bulkedge} in the context of covariant operators was proved in \cite{SKR,KRS} where the l.h.s. also contains a disorder average. It is shown how this result can be recovered in Section~\ref{sec-covariant}. The pointwise equality \eqref{eq-bulkedgepointwise} was first proved in \cite{EG}. The rough idea for the proof given below is due to Macris \cite{Mac}, but as already indicated in the introduction his argument contained several imprecisions which are corrected below. 

\vspace{.2cm}

\begin{figure}
\begin{center}
\includegraphics[height=7cm]{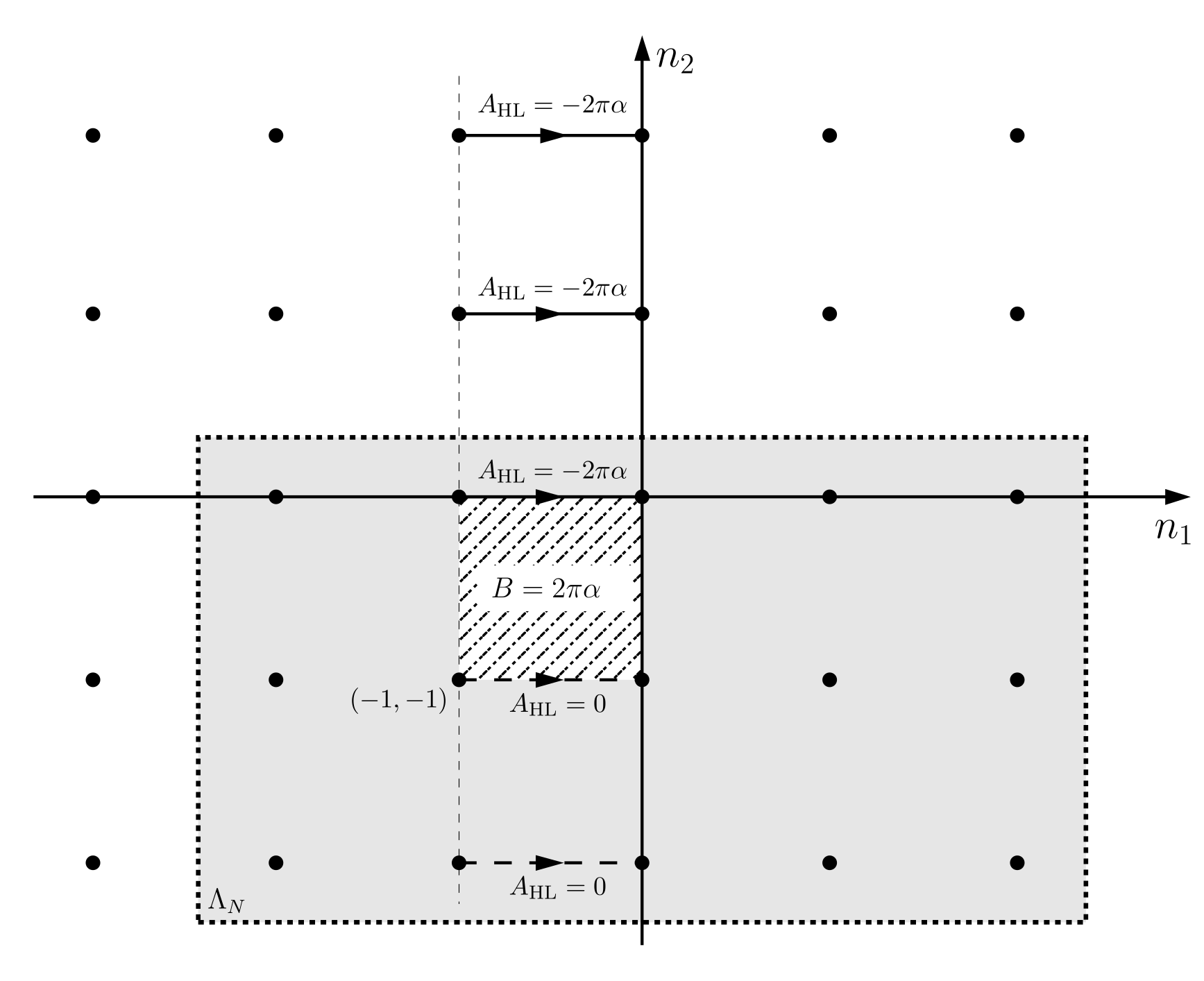}
\caption{\it 
Illustration of the geometry in the proof of {\rm Theorem~\ref{theo-bulkedge}}.
\label{fig-bulkedge}
}
\end{center}
\end{figure}

\noindent {\bf Proof} of Theorem~\ref{theo-bulkedge}. The proof begins by exploiting that \eqref{eq-LaughlinFlowFiniteVol} holds for all Hamiltonians within the class \eqref{eq-disorderedHarper} provided the Gap hypothesis holds, and also for all choices of the gauge. Here we choose $\Lambda_N=\{(n_1,n_2)\,|\,|n_1|\leq N\,,\;\;-N\leq n_2\leq 0\}$ and the gauge \eqref{eq-gaugechoice} for a flux $\alpha$ attached to $m=(m_1,m_2)=(-1,-1)$, see the illustration in Figure~\ref{fig-bulkedge}. Hence the magnetic translations $\widetilde{S}^{B,\alpha}_j$ defined in \eqref{eq-gaugechoice} are used in \eqref{eq-disorderedHarper} so that the Hamiltonian will be 
$$
\widetilde{H}_\alpha
\;=\;
\sum_{|n_1|\leq R,\,|n_2|\leq R}\,  t_{n}(\alpha)\;(\widetilde{S}_1^{B,\alpha})^{n_1}\;(\widetilde{S}_2^{B,\alpha})^{n_2}
\;+\;
V
\;,
$$
and the restriction to $\Lambda^c_N=\ZM^2\setminus \Lambda_N$ is as above
$$
\widetilde{H}_\alpha^N
\;=\;
\Pi_{\Lambda^c_N}\,\widetilde{H}_\alpha\,\Pi_{\Lambda^c_N}
$$
The tilde and $N$ on the Hamiltonian $\widetilde{H}_\alpha^N$ indicate the choice of half-line gauge and the dependence on  physical space. With these notations, \eqref{eq-LaughlinFlowFiniteVol} becomes
\begin{equation}
\label{eq-LaughlinFlow3}
\Ind(P_\mu FP_\mu)
\;=\;
-\;\int_0^1 d\alpha\;
\Tr\bigl(
g'(\widetilde{H}_\alpha^N)\,\partial_\alpha \widetilde{H}_\alpha^N
\bigr)
\;.
\end{equation}
Next let us set $\Pi_r=\sum_{n_2\geq 0}|r,n_2\rangle\langle r,n_2|$ and $\Pi_{[-R,R]}=\sum_{r=-R}^R\,\Pi_r$. With these notations, the formula after \eqref{eq-gaugechoice} reads $\widetilde{S}^{B,\alpha}_1=e^{-\imath BX_2}S_1(\one+(e^{2\pi\imath\alpha}-1)\Pi_0)$ and it follows
$$
\partial_\alpha\widetilde{S}^{B,\alpha}_1
\;=\;
2\pi \imath\,e^{2\pi\imath\alpha}\,e^{-\imath BX_2} \,S_1\,\Pi_0
\;=\;
2\pi \imath\,\widetilde{S}^{B,\alpha}_1 \Pi_0
\;,
\qquad
\partial_\alpha\widetilde{S}^{B,\alpha}_2
\;=\;
0
\;.
$$
As $S_1\Pi_0=S_1(S_1^*\Pi_<S_1-\Pi_<)=[\Pi_<,S_1]$ and $\Pi_<$ commutes with any multiplication operator, one obtains $\widetilde{S}^{B,\alpha}_1\Pi_0=[\Pi_<,\widetilde{S}^{B,\alpha}_1]$ so that
\begin{equation}
\label{eq-DerivId}
\partial_\alpha\widetilde{S}^{B,\alpha}_1
\;=\;
2\pi \imath\,[\Pi_<,\widetilde{S}^{B,\alpha}_1]
\;,
\qquad
\partial_\alpha\widetilde{S}^{B,\alpha}_2
\;=\;
0
\;.
\end{equation}
As $\Pi_<$ commutes with $\widetilde{S}^{B,\alpha}_2$ as well as any potential, one concludes using the derivative properties of both sides of the equality (Leibniz rule) that
$$
\partial_\alpha \widetilde{H}_\alpha^N
\;=\;
2\pi\imath\,[\Pi_<,\widetilde{H}_\alpha^N]
\;.
$$
As the sum in the Hamiltonian $\widetilde{H}_\alpha^N$ is restricted to $|n|\leq R$ and $[\Pi_<,\widetilde{S}^{B,\alpha}_1]=\Pi_{-1}\widetilde{S}^{B,\alpha}_1\Pi_0$ leading to a similar identity for powers of $\widetilde{S}^{B,\alpha}_1$, one has 
$$
\partial_\alpha \widetilde{H}_\alpha^N
\;=\;
\Pi_{[-R,R]}\,\partial_\alpha\widetilde{H}_\alpha^N\,\Pi_{[-R,R]}
\;.
$$
Furthermore, let $\widehat{H}_\alpha=\widetilde{H}_\alpha^\infty$ be the half-space restriction of $H$ given by \eqref{eq-disorderedHarper} with the magnetic translations $\widetilde{S}^{B,\alpha}_j$. Then, for $N\geq R$, one furthermore has $\partial_\alpha \widetilde{H}_\alpha^N=\Pi_{[-R,R]}\,\partial_\alpha\widehat{H}_\alpha\,\Pi_{[-R,R]}$ and similarly for the commutators with $\Pi_<$. Hence \eqref{eq-LaughlinFlow3} becomes
\begin{equation}
\label{eq-fluxhalfplane}
\Ind(P_\mu FP_\mu)
\;=\;
-\;2\pi\,\int_0^1 d\alpha\;
\Tr\bigl(\Pi_{[-R,R]}\,
g'(\widetilde{H}_\alpha^N)\,\Pi_{[-R,R]}\,\imath\,[\Pi_<,\widehat{H}_\alpha]
\bigr)
\;.
\end{equation}
Now Lemma~\ref{lem-matrixelementbound} below shows that $\Pi_{[-R,R]} \,g'(\widetilde{H}_\alpha^N)\,\Pi_{[-R,R]}$ is trace-class uniformly in $N$ and that $\widetilde{H}_\alpha^N$ can be replaced by $\widehat{H}_\alpha$. Hence, taking the limit $N\to\infty$  of \eqref{eq-fluxhalfplane} shows
$$
\Ind(P_\mu FP_\mu)
\;=\;
-\;2\pi\;\int_0^1 d\alpha\;
\Tr\bigl(\Pi_{[-R,R]}\,
g'(\widehat{H}_\alpha)\,\Pi_{[-R,R]}\,\imath\,[\Pi_<,\widehat{H}_\alpha]
\bigr)
\;.
$$
Finally the projections $\Pi_{[-R,R]}$ can be dropped again because they are contained in $\imath\,[\Pi_<,\widehat{H}_\alpha]$ anyway. Now the Hamiltonian  $\widehat{H}_\alpha$ does depend on $\alpha$ through the gauge for the flux quantum at $(-1,-1)$, but the half-line gauge on the upper half-plane $\ZM\times\NM$ does not lead to a magnetic field there. Thus there exists a gauge transformation from $\widehat{H}_\alpha$ to $\widehat{H}$, the latter of which has only a gauge potential for the constant magnetic field ({\it e.g.} the flux $\alpha$ can be realized by a half-line type potential in the lower half-plane). As again the trace on the r.h.s. is invariant under gauge transformation one may set $\alpha=0$. This completes the proof of Theorem~\ref{theo-bulkedge}. 
\hfill $\Box$

\begin{lemma} 
\label{lem-matrixelementbound} 
$\Pi_{[-R,R]}\,g'(\widetilde{H}_\alpha^N)\,\Pi_{[-R,R]}$ converges as $N\to\infty$ to $\Pi_{[-R,R]}\,
g'(\widehat{H}_\alpha)\,\Pi_{[-R,R]}$ in trace norm.
\end{lemma}

\noindent {\bf Proof.} First let us recall the argument from the appendix of \cite{EG}, see also Section~5 in \cite{Sch0}, showing that $\Pi_{[-R,R]}\, g'(\widehat{H}_\alpha)\,\Pi_{[-R,R]}$ is traceclass. By the Helffer-Sj\"orstrand formula:
$$
g'(\widehat{H}_\alpha)
\;=\;
\int \frac{dz}{2\pi}
\,
\widetilde{g}(z,\overline{z})
\,
(z-\widehat{H}_\alpha)^{-1}
\;.
$$
Here $\widetilde{g}$ is the derivative of an adequate quasi-analytic extension of $g'$ and the two-dimensional integral is over a rectangle in the complex plane the $x$-axis of which is contained in the support of $g'$. In particular, $\widetilde{g}$ decreases arbitrarily fast on the real axis. By a Combes-Thomas estimate the resolvent decays exponentially and this allows to conclude by the arguments in \cite{EG,Sch0}. Furthermore,
$$
g'(\widetilde{H}_\alpha^N)
-
g'(\widehat{H}_\alpha)
\;=\;
\int \frac{dz}{2\pi}
\,
\widetilde{g}(z,\overline{z})
\,
(z-\widetilde{H}_\alpha^N)^{-1}
\bigl(
\widetilde{H}_\alpha^N
-
\widehat{H}_\alpha
\bigr)
(z-\widehat{H}_\alpha)^{-1}
\;.
$$
The crucial fact is now that $\widetilde{H}_\alpha^N-\widehat{H}_\alpha$ has vanishing matrix elements only in the lower half-plane $\ZM\times(-\NM)$ and outside of $\Lambda_N$. Therefore again a Combes-Thomas estimate, now for both resolvents, allows to conclude.
\hfill $\Box$

\section{Covariant Hamiltonians}
\label{sec-covariant}

Let $\Omega$ be a compact space furnished with an action $T$ of $\ZM^2$ and an invariant and ergodic probability measure $\PP$. A family of operators $A=(A_\omega)_{\omega\in\Omega}$ on $\ell^2(\ZM^2)$ is called covariant w.r.t. the magnetic translations $S^B_1$, $S^B_2$ if and only if
$$
(S^B_j)^*\,A_\omega\, S^B_j
\;=\;
A_{T_j\omega}
\;,
$$
where $T_j=T_{e_j}$ for $j=1,2$ are the two generators of the actions of $\ZM$. The finite range covariant operators form an algebra and its closure is a C$^*$-algebra $\Aa$ which has the structure of a crossed product algebra \cite{BES} and its von Neumann closure w.r..t. $\|A\|=\PP\mbox{-esssup}\|A_\omega\|$ is denoted by $L^\infty(\Aa,\PP)$.  For a covariant projection $P=(P_\omega)_{\omega\in\Omega}$ with sufficient decay of matrix elements off the diagonal (so that, in particular, the commutators $[X_j,P_\omega]$ are bounded), the Chern number is defined by \cite{BES}
$$
\mbox{\rm Ch}(P)
\;=\;
2\pi\imath\;\int \PP(d\omega)\;\langle 0|P_{\omega}[[X_1,P_{\omega}],[X_2,P_{\omega}]]|0\rangle
\;.
$$
One of the main results of \cite{BES} is that, with $F$ defined in \eqref{eq-Fphase}, the operators $P_\omega FP_\omega$ are almost surely Fredholm operators on $P_\omega\ell^2(\ZM^2)$ and the indices are $\PP$-almost surely constant and given by
$$
\mbox{\rm Ch}(P)
\;=\;
\mbox{\rm Ind}(P_\omega FP_\omega)
\;,
$$
Furthermore, if $P$ is the Fermi projection, then $\mbox{\rm Ch}(P)$ is equal the zero temperature Hall conductance as given by the Kubo formula.

\vspace{.2cm}

Now the equality \eqref{eq-bulkedgepointwise} connecting the index to the boundary current density holds pointwise, that is for every realization $\omega$. The following theorem is the version of the bulk-edge correspondence proved in \cite{SKR,KRS}.

\begin{theo} 
\label{theo-bulkedgeaveraged} 
Suppose that the closed set $\Delta\subset\RM$ lies in an almost sure gap of a covariant family  $H=(H_\omega)_{\omega\in\Omega}$ of Hamiltonians of the form {\rm \eqref{eq-disorderedHarper}}.  Further let $g:\RM\to[0,1]$ be any smooth non-increasing function which is equal to $1$ on the left of $\Delta$ and $0$ on the right of $\Delta$. Then
\begin{equation}
\label{eq-bulkedgeaveraged}
\sum_{n_2\geq 0}\, \int \PP(d\omega)\,
\langle 0,n_2| g'(\widehat{H}_\omega)\,\imath \,[X_1,\widehat{H}_\omega]
|0,n_2\rangle
\;=\;
-\;\frac{1}{2\pi}\;\Ind(P_{\mu,\omega} FP_{\mu,\omega})
\;,
\end{equation}
where $P_{\mu,\omega}=\chi(H_\omega\leq \mu)$ is the Fermi projection on energies below $\mu\in\Delta$ and on the r.h.s. the almost sure index is taken.
\end{theo}

\noindent {\bf Proof.} Let us begin by extending the notation  \eqref{eq-quarterplane} a bit by setting
$$
\Pi_{<n_1}
\;=\;
\sum_{n'_1<n_1}\;\sum_{n_2>0}\;|n'_1,n_2\rangle\langle n'_1,n_2|
\;,
\qquad
\Pi_{N}
\;=\;
\sum_{|n_1|\leq N}\;\sum_{n_2>0}\;|n_1,n_2\rangle\langle n_1,n_2|
\;.
$$
With these notations, one has
\begin{align*}
\Tr\bigl(g'(\widehat{H}_\omega)\,\imath \,[\Pi_<,\widehat{H}_\omega]\bigr)
& =\;
\Tr\bigl(g'(\widehat{H}_{T_1^{n_1}\omega})\,\imath \,[\Pi_{<n_1},\widehat{H}_{T_1^{n_1}\omega}]\bigr)
\\
& =\;
\Tr\bigl(\Pi_N\,g'(\widehat{H}_{T_1^{n_1}\omega})\,\imath \,[\Pi_{<n_1},\widehat{H}_{T_1^{n_1}\omega}]\,\Pi_N\bigr)
\;,
\end{align*}
for $N>|n_1|+R$ where $R$ is the range of $H_\omega$. Now let $\EE$ denote the average w.r.t. $\PP$. Using the invariance of $\PP$ it follows that
$$
\EE\;
\Tr\bigl(g'(\widehat{H}_\omega)\,\imath \,[\Pi_<,\widehat{H}_\omega]\bigr)
\;=\;
\EE\;\frac{1}{2N+1}\;
\Tr\Big(\Pi_{N+R}\,g'(\widehat{H}_{\omega})\,\imath \,\Big[\sum_{|n_1|\leq N}\Pi_{<n_1},\widehat{H}_{\omega}\Big]\,\Pi_{N+R}\Big)
\;.
$$
Now $\Pi_{N+R}\,g'(\widehat{H}_{\omega})$ is traceclass, and 
$$
\mbox{\rm w -}\!\!\lim_{N\to\infty}
\;\Big[\sum_{|n_1|\leq N}\Pi_{<n_1},\widehat{H}_{\omega}\Big]
\;=\;[X_1,\widehat{H}_{\omega}]
\;.
$$
Furthermore, by Birkhoff's theorem for any covariant operator family $A=(A_\omega)_{\omega\in\Omega}$ in the $1$-direction, summable in the $2$-direction, one has
$$
\lim_{N\to\infty}
\frac{1}{2N+1}\;
\Tr\Big(\Pi_{N+R}\,A_\omega\,\Pi_{N+R}\Big)
\;=\;
\sum_{n_2\geq 0}\, \int \PP(d\omega)\,
\langle 0,n_2| A_\omega
|0,n_2\rangle
\;.
$$
Combining these facts concludes the proof.
\hfill $\Box$

\section{Spectral flows in presence of symmetries}
\label{sec-symmetries}

In this section the fate of the flux tube argument in presence of fundamental discrete symmetries SLS, TRS and PHS is analyzed. The implementation of these symmetries in two-dimensional tight-binding models and some basic consequences are discussed in Section~\ref{sec-symrev}. In the following sections, a flux tube is inserted into such systems and this breaks these symmetries. Nevertheless the associated spectral flow has signatures which are characteristic for the underlying symmetry.

\subsection{Review of fundamental discrete symmetries}
\label{sec-symrev}

For the implementations of the symmetries, one supposes given (one of more of) three commuting unitaries $I_{\mbox{\rm\tiny tr}}$, $\KPH$ and $\KSL$ with real $I_{\mbox{\rm\tiny tr}}$ and $\KPH$ satisfying $(I_{\mbox{\rm\tiny tr}})^2=\eta_{\mbox{\rm\tiny tr}}\one$ and $(\KPH)^2=\eta_{\mbox{\rm\tiny ph}}\one$ with $\eta_{\mbox{\rm\tiny tr}},\eta_{\mbox{\rm\tiny ph}}\in\{-1,1\}$, and furthermore $(\KSL)^2=\one$. Then the Hamiltonian is said to have respectively TRS, PHS, SLS if
\begin{align}
& (I_{\mbox{\rm\tiny tr}})^*\,\overline{H}\,I_{\mbox{\rm\tiny tr}}\;=\;H
\;,
\label{eq-TRS}
\\
& (\KPH)^*\,\overline{H}\,\KPH\;=\;-\,H
\;,
\label{eq-PHS}
\\
& (\KSL)^*\,{H}\,\KSL\,=\,-\,H
\;.
\label{eq-SLS}
\end{align}
Here $\overline{H}$ denotes the complex conjugate of $H$ associated to a given real structure on the Hilbert space. The TRS and PHS are said to be even or odd pending on the signs  in $\eta_{\mbox{\rm\tiny tr}}$ and $\eta_{\mbox{\rm\tiny ph}}$. As the SLS does not involve a complex conjugate it is not necessary to consider the case of even and odd SLS (because one can remove the sign by considering $\imath\,\KSL$ instead of $\KSL$). Let us also point out that, if both TRS and PHS are given, then one has a SLS by setting $\KSL=I_{\mbox{\rm\tiny tr}}\KPH$ or  $\KSL=\imath\,I_{\mbox{\rm\tiny tr}}\KPH$ pending on the signs. Therefore, it is possible to obtain the following combinations of TRS, PHS and SLS \cite{SRFL}: no symmetry, only SLS, only PHS (2 cases), only TRS (2 cases), both PHS and TRS (4 cases). In total there are therefore 10 classes which are listed in Table~\ref{table1}. Following \cite{Kit}, the 10 classes are separated in one group (A and AIII) of cases which do not use complex conjugation, and the remaining $8$ which do. These latter $8$ are ordered according to the $K$-theoretic considerations given in the introduction.

\vspace{.2cm}

\noindent {\bf Example} In two-dimensional tight-binding models the symmetries are typically implemented in the enlarged Hilbert space $\Hh=\ell^2(\ZM^2)\otimes\CM^L$ where the finite dimensional fiber $\CM^L$ allows to describe further degrees of freedom. The most general fiber  is of the form $\CM^L=\CM^{2s+1}\otimes\CM^2_{\mbox{\rm\tiny ph}}\otimes \CM^2_{\mbox{\rm\tiny sl}}\otimes \CM^N$ where $\CM^{2s+1}$ is associated to the spin degree of freedom of a spin $s\in\frac{1}{2}\,\NM_0$, $\CM^2_{\mbox{\rm\tiny ph}}$ and $\CM^2_{\mbox{\rm\tiny sl}}$ are the particle-hole and sublattice degrees of freedom, and $\CM^N$ describes any further internal degrees of freedom over each site of the lattice $\ZM^2$, like larger elementary cells or possibly several orbitals. It is, however, also possible that $\CM^L$ only contains fewer factors, say only one of them. On the fibers $\CM^{2s+1}$, $\CM^2_{\mbox{\rm\tiny ph}}$ and $\CM^2_{\mbox{\rm\tiny sl}}$ now act unitary matrices which naturally extend to $\Hh$ by tensorizing with the identity. On the spin component acts the rotation $I_{\mbox{\rm\tiny tr}}$ in spin space by $180$ degrees. Let $s^y$ be a purely imaginary irreducible representation of the $y$-component of the spin on $\CM^{2s+1}$. Then a real unitary is
$$
I_{\mbox{\rm\tiny tr}}\;=\;e^{\imath\pi s^y}
\;,
\qquad
(I_{\mbox{\rm\tiny tr}})^2\;=\;
(-1)^{2s}\,\one
\;.
$$
On the particle-hole and sublattice fibers act
\begin{equation}
\label{eq-PHSSLS}
\KPH
\;=\;
\begin{pmatrix}
0 & \eta_{\mbox{\rm\tiny ph}}  \\ 1 & \;\; 0 
\end{pmatrix}
\;,
\qquad
\KSL
\;=\;
\begin{pmatrix}
1 & \;\;0 \\ 0 & -1 
\end{pmatrix}
\;,
\qquad
\KPH^2\;=\;\eta_{\mbox{\rm\tiny ph}} \one
\;,\;\;
\KSL^2\;=\;\one
\;.
\end{equation}
In concrete models, the symmetries appear naturally. Of course, the above representation is not unique and it may be better to work in different one.
\hfill $\diamond$

\vspace{.2cm}

\noindent {\bf Remark 1} 
By going into the spectral representation of $\KPH$, it is always possible by an orthogonal change of basis  to find a grading of the Hilbert space $\Hh$ such that $\KPH$ is of the form given in \eqref{eq-PHSSLS}. Then $\Hh=\Hh'\otimes \CM^2_{\mbox{\rm\tiny ph}}$ and the PHS \eqref{eq-PHS} is equivalent to the Hamiltonian being of the following form in particle-hole grading:
\begin{equation}
H
\;=\;
\begin{pmatrix}
h & \Delta \\
-\,\eta_{\mbox{\rm\tiny ph}}\,\overline{\Delta} & -\overline{h}
\end{pmatrix}
\;.
\label{eq-BdG}
\end{equation}
This is the conventional Bogoliubov-de Gennes (BdG) form of the Hartree-Fock approximation to BCS models and $\Delta$ is then called the pair creation potential. The even PHS $\eta_{\mbox{\rm\tiny ph}}=1$ covers BdG operators having no further symmetry, whereas the odd PHS $\eta_{\mbox{\rm\tiny ph}}=-1$ appears for reduced operators for fully spin-rotation invariant systems \cite{AZ}. In \cite{DDS} there is a list of standard BdG models, and the topologically non-trivial ones are used as examples below. 
\hfill $\diamond$

\vspace{.2cm}

\noindent {\bf Remark 2} 
It is often useful to pass to the so-called Majorana representation obtained by doing a Cayley transformation in the particle-hole space
\begin{equation}
H_\Maj
\;=\;
C^t\,H\,\overline{C}
\;,
\qquad
C\;=\;
\frac{1}{\sqrt{2}}
\begin{pmatrix}
\one & -\,\imath\,\one \\
\one & \imath\,\one
\end{pmatrix}
\;.
\label{eq-BdGMaj}
\end{equation}
Then the particle-hole symmetry \eqref{eq-PHS} becomes $(C^*\KPH \overline{C})^*\overline{H_\Maj}  (C^*\KPH \overline{C})=-  H_\Maj$. For $\eta_{\mbox{\rm\tiny ph}}=1$, one finds $C^*\KPH \overline{C}=\one$ and thus
$$
H_\Maj
\;=\;
H_\Maj^*
\;=\;
-\,\overline{H_\Maj}
\;=\;
-\,H_\Maj^t
\;,
$$
namely $H_\Maj$ is purely imaginary and antisymmetric. More explicitly, in terms of the matrix entries of \eqref{eq-BdG} for $\eta_{\mbox{\rm\tiny ph}}=1$
\begin{equation}
\label{eq-MajRep}
H_\Maj
\;=\;
\imath\,
\begin{pmatrix}
\mbox{\rm im}(h)+\mbox{\rm im}(\Delta) & \mbox{\rm re}(h)-\mbox{\rm re}(\Delta) \\
-\mbox{\rm re}(h)-\mbox{\rm re}(\Delta) & \mbox{\rm im}(h)-\mbox{\rm im}(\Delta)
\end{pmatrix}
\;,
\end{equation}
with a real symmetric operator $\mbox{\rm re}(h)=\frac{1}{2}(h+\overline{h})$ and real skew-symmetric operators $\mbox{\rm im}(h)=\frac{1}{2\imath}(h-\overline{h})$, $\mbox{\rm re}(\Delta)=\frac{1}{2}(\Delta+\overline{\Delta})$ and $\mbox{\rm im}(\Delta)=\frac{1}{2\imath}(\Delta-\overline{\Delta})$. 
\hfill $\diamond$

\vspace{.2cm}

Now let us collect a few basic spectral implications of the various symmetries. Both the SLS and the PHS (even or odd) imply that the spectrum of the Hamiltonian satisfies the reflection property $\sigma(H)=-\sigma(H)$. Less well-known, but equally elementary, is that the essentially gapped BdG models fall into two classes. This requires no spacial structure of the Hamiltonian and is merely related to the PHS. The argument leading to the $\ZM_2$ invariant below is essentially the same as the one used by Atiyah-Singer to introduce a $\ZM_2$ index for real skew-symmetric operators \cite{AS}. 

\begin{proposi} 
\label{prop-BdGclass} 
The {\rm BdG} Hamiltonians with $0\not\in\sigma_\ess(H)$ and even {\rm PHS} fall into two classes which are labelled by $\Ind_2(H)=\dim (\Ker(H))\;\mbox{\rm mod}\,2\in\ZM_2$, which cannot be homotopically connected without closing the central gap.
\end{proposi}

\noindent {\bf Proof.} The spectrum of the Hamiltonian satisfies $\sigma(H)=-\sigma(H)\subset\RM$ and the gap hypothesis implies that $H$ is a Fredholm operator. Eigenvalues come in pairs $E,-E$ which may merge under a homotopy into $0$ and thus change the dimension of the kernel by $2$. However, under any homotopy $\Ind_2(H)$ is invariant as claimed. 
\hfill $\Box$

\vspace{.2cm}

The argument of Proposition~\ref{prop-BdGclass} also applies to an odd PHS, but then the kernel of $H$ is always even dimensional by an argument similar to the following  well-known Kramers' degeneracy. 

\begin{proposi} 
\label{prop-Kramersdeg} 
Suppose $H$ has odd {\rm TRS}. Then any discrete eigenvalue is even multiplicity and the eigenspace is left invariant under $\psi\mapsto I_{\mbox{\rm\tiny tr}}\overline{\psi}$.
\end{proposi}

\noindent {\bf Proof.} If $H\psi=E\psi$, then also $HI_{\mbox{\rm\tiny tr}}\overline{\psi}=E\,I_{\mbox{\rm\tiny tr}}\overline{\psi}$. Let us show that the vectors $\psi$ and $I_{\mbox{\rm\tiny tr}}\overline{\psi}$ are linear independent. Indeed, suppose $\psi=\lambda I_{\mbox{\rm\tiny tr}}\overline{\psi}$ for some $\lambda\in\CM$. Then $\psi=\lambda\,I_{\mbox{\rm\tiny tr}}\overline{\lambda}I_{\mbox{\rm\tiny tr}}{\psi}=-|\lambda|^2\psi$ which implies $\lambda=0$. This argument can be extended to deal with higher dimensional eigenspaces.
\hfill $\Box$

\subsection{Particle-hole symmetric systems}
\label{sec-PHS}

In this section, the Hilbert space is $\Hh=\ell^2(\ZM^2)\otimes\CM^N\otimes\CM^2_{\mbox{\rm\tiny ph}}$ and $\KPH$ is of the form \eqref{eq-PHSSLS}. Thus one has the BdG representation \eqref{eq-BdG}, and the operators $h=h^*$ and $\Delta=-\eta_{\mbox{\rm\tiny ph}} \Delta^t$ therein act on $\ell^2(\ZM^2)\otimes \CM^N$. Now a family $(H_\alpha)_{\alpha\in[0,1]}$ of compact perturbations of $H=H_0$ of the form
\begin{equation}
H_\alpha
\;=\;
\begin{pmatrix}
h_\alpha & \Delta_\alpha \\
-\,\eta_{\mbox{\rm\tiny ph}}\,\overline{\Delta_{-\alpha}} & -\overline{h_{-\alpha}}
\end{pmatrix}
\;.
\label{eq-BdGalpha}
\end{equation}
will be considered which is, moreover, supposed to satisfy
\begin{equation}
F\,H_\alpha \,F^*\;=\;H_{\alpha+1}\;.
\label{eq-BdGalpha2}
\end{equation}
where again $F=F\otimes \one$ denotes the natural extension from $\ell^2(\ZM^2)$ to $\Hh$. Further below several models and physical contexts will be discussed that lead to such families of Hamiltonians and, of course, the $\alpha$ then also corresponds to flux tubes. Let us now first go on with the analysis of such a family $(H_\alpha)_{\alpha\in[0,1]}$ of operators. The form of \eqref{eq-BdGalpha} combined with \eqref{eq-BdGalpha2} implies
$$
\KPH^*\,\overline{H_\alpha}\,\KPH\;=\;-\,H_{-\alpha}
\;=\;
-\,F^*\,H_{1-\alpha}\,F
\;,
$$
so that the spectra satisfy
\begin{equation}
\label{eq:spec_sym_BdG}
\sigma(H_\alpha)
\;=\;
-\,\sigma(H_{-\alpha})
\;=\;
-\,\sigma(H_{1-\alpha})
\;.
\end{equation}
These identities were already used in \cite{Roy,EF} as the basic tool to study the spectral flow of operators with PHS. For $\alpha=0$ and $\alpha=\frac{1}{2}$, this is the well-known spectral symmetry of BdG operators. Let us point out that for $\alpha\not=0,\frac{1}{2},1$, the operator $H_\alpha$ does not have PHS, and for $\alpha=\frac{1}{2}$ there is a modified PHS with a unitary which is neither real nor does it square to a multiple of the identity. Nevertheless, the argument of Proposition~\ref{prop-BdGclass} applies so that $\mbox{\rm Ind}_2(H_{\frac{1}{2}})$ is well-defined.

\begin{theo} 
\label{theo-PHS} 
Let $H=H_0$ be a two-dimensional {\rm BdG} Hamiltonian with either even or odd {\rm PHS}. Suppose that $0$ lies in a gap of $H$  and let $P=\chi(H\leq 0)$ be the Fermi projection. Let $H_{\frac{1}{2}}$ the Hamiltonian obtained by setting $\alpha=\frac{1}{2}$ in a family of the form {\rm \eqref{eq-BdGalpha}} and {\rm \eqref{eq-BdGalpha2}}. Then
$$
\mbox{\rm Ind}(PFP)\;\mbox{\rm mod}\;2
\;=\;
\mbox{\rm Ind}_2(H_{\frac{1}{2}})
\;.
$$
In particular, if $\mbox{\rm Ind}(PFP)$ is odd, then the half-flux operator $H_{\frac{1}{2}}$ has at least one zero energy eigenvalue. 
\end{theo}

\begin{figure}
\begin{center}
\includegraphics[height=6cm]{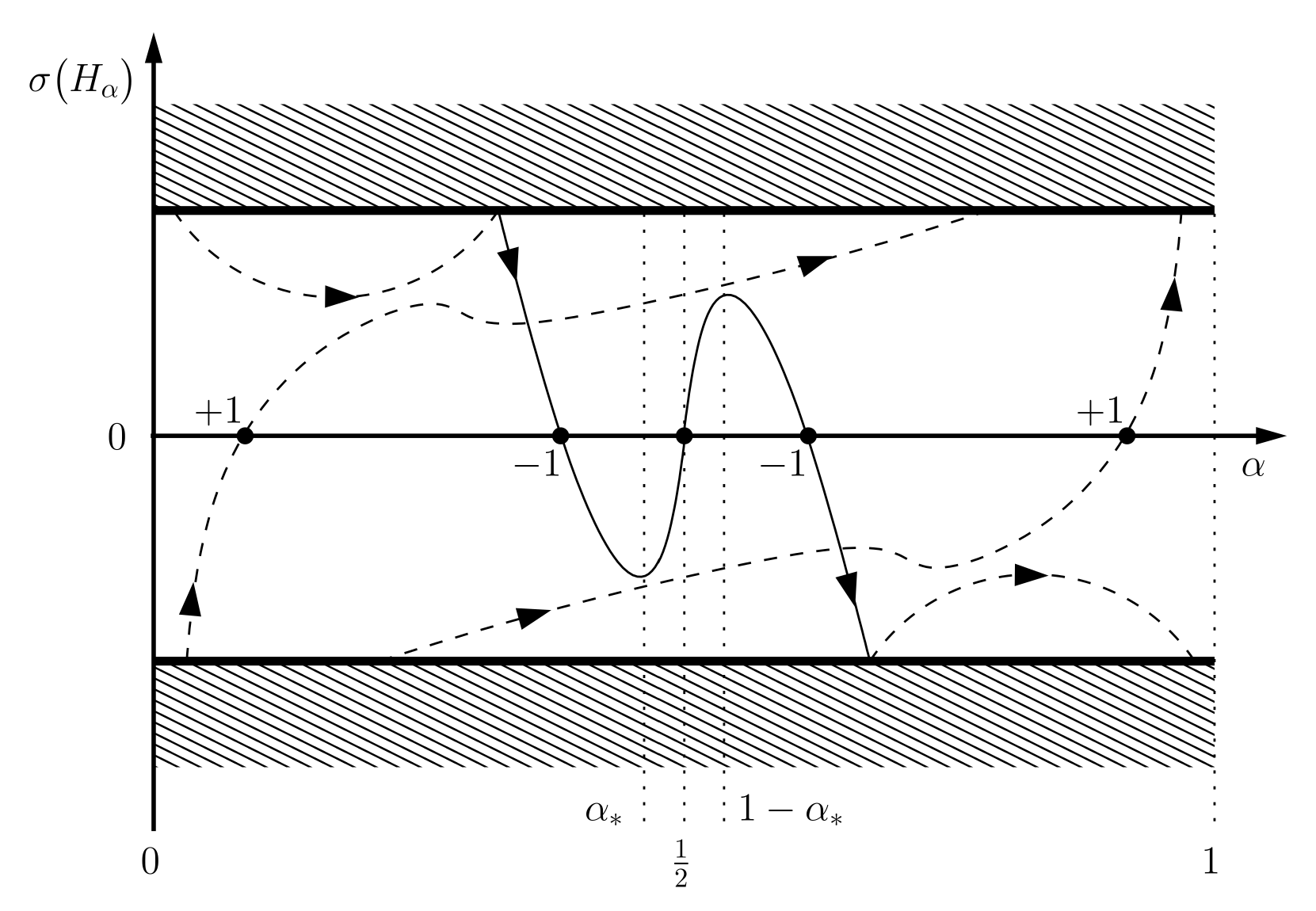}
\caption{\it 
Typical pattern of the spectral flow associated with a flux tube insertion $\alpha\mapsto H_\alpha$ subjected to a symmetry of type  \eqref{eq:spec_sym_BdG}.
\label{fig-BdG}
}
\end{center}
\end{figure}

\noindent {\bf Proof.}
The relation \eqref{eq:spec_sym_BdG} for the spectrum of $H_\alpha$ implies
$$
\SF\bigl(\alpha\in[0,\alpha_\ast]\mapsto H_{\alpha} \;\mbox{\rm by}\;0\bigr)\;=\;\SF\bigl(\alpha\in[1-\alpha_\ast,1]\mapsto H_{\alpha}\;\mbox{\rm by}\;0\bigr)\;=\;N
\;,
$$
for all $0<\alpha_\ast<\frac{1}{2}$ where $N$ is some integer depending on $\alpha_\ast$. This equality is evident from \eqref{eq:spec_sym_BdG}, see also Figure~\ref{fig-BdG}. Then one has 
$$
\mbox{\rm Ind}(PFP)\;=\;\SF\bigl(\alpha\in[0,1]\mapsto H_{\alpha} \;\mbox{\rm by}\;0\bigr)\;=\; 2N\;+\; \SF\bigl(\alpha\in[\alpha_\ast,1-\alpha_\ast]\mapsto H_{\alpha} \;\mbox{\rm by}\;0\bigr)
\;.
$$
Now the choice of $\alpha_\ast$ is arbitrary and it is possible to conider the limit $\alpha_*\uparrow\frac{1}{2}$. This implies that an odd value of $\mbox{\rm Ind}(PFP)$ is possible if and only if $0$ is an eigenvalue of odd multiplicity for $H_{\frac{1}{2}}$, see again Figure~\ref{fig-BdG} for an illustration.
\hfill $\Box$

\vspace{.2cm}

\noindent {\bf Remark 1} Theorem~\ref{theo-PHS} is a pointwise statement for a given fixed BdG operator $H$. If one has a covariant family of BdG Hamiltonians, the indices $\mbox{\rm Ind}(PFP)$ are almost surely equal to the Chern number $\mbox{\rm Ch}(P)$ and the hypothesis of the theorem is on the parity of the Chern number, just as in \cite{Roy,EF}. 
\hfill $\diamond$

\vspace{.2cm}

\noindent {\bf Remark 2} One may be tempted to apply Theorem~\ref{theo-PHS} to $H_\alpha=F^{\alpha} H_0 F^{-\alpha}$ with an $H_0$ having BdG symmetry and an odd index. However, clearly there is no spectral flow and no zero mode for $H_{\frac{1}{2}}$. Indeed, this is no contradiction because $H_\alpha-H_0$ is not compact.
\hfill $\diamond$

\vspace{.2cm}

\noindent {\bf Example 1} The importance of the above theorem is rooted in the influential paper of Read and Green \cite{RG}. It claims that when a vortex solution of the Landau-Ginzburg equation for the pair creation potential $\Delta$ is used in $p+i p$ superconductor, then the associated BdG operator has a zero energy mode leading to a Majorana fermion in the second quantized representation. Theorem~\ref{theo-PHS} provides a rigorous proof for the existence of such zero energy modes for a wide class of tight-binding models.  Modeling a vortex of $\Delta$ in a tight-binding model is a delicate issue, see the discussions in \cite{VMFT}. Here we choose the operators $h_\alpha$ and $\Delta_\alpha$ in \eqref{eq-BdGalpha} to be
\begin{equation}
\label{eq-BdGentries}
h_\alpha
\; =\;
\sum_{|n|\leq R}\,  t_{n}(\alpha)\;(S_1^{B,\alpha})^{n_1}(S_2^{B,\alpha})^{n_2}
\;+\;
V
\;,
\qquad
\Delta_\alpha
\;=\;
\sum_{|n|\leq R}\, d_{n}(\alpha)\;(S^{0,\alpha}_1)^{n_1}(S^{0,\alpha}_2)^{n_2}\;+\;W
\;,
\end{equation}
where $t_n(\alpha),d_n(\alpha)\in\Kk^\sim\otimes\mbox{\rm Mat}(N\times N,\CM)$ with $t_n(\alpha)$ satisfying \eqref{eq-hoppingcond} and $d_n(\alpha)$
$$
d_{n}(\alpha)\;=\;-\,\eta_{\mbox{\rm\tiny ph}}
(S_2^{0,\alpha})^{n_2}\; (S_1^{0,\alpha})^{n_1}\;d_n(-\alpha)^*
\;(S_2^{0,\alpha})^{n_2}\;(S_1^{0,\alpha})^{n_1}\;,
\qquad
F d_n(\alpha) F^*\;=\;d_n(\alpha+1)
\;,
$$
and $V$ and $W$ being matrix-valued potentials, {\it e.g.} $W=\sum_{n\in\ZM^2}w_n\,|n\rangle\langle n|$. These conditions assure that $H_\alpha$ is self-adjoint and that \eqref{eq-BdGalpha2} holds. Furthermore, the compacticity of $H_\alpha-H_0$ follows from Proposition~\ref{prop-magtransprop}. Hence Theorem~\ref{theo-PHS} implies the existence of a zero mode for $H_{\frac{1}{2}}$ provided $\mbox{\rm Ind}(PFP)$ is odd. 

\vspace{.1cm}

A concrete model for which all this is satisfied is a tight-binding $p+ip$ wave superconductor with $h=S_1+S_1^*+S_2+S_2^*-\mu$ and $\Delta=\delta(S_1-S_1^*+\imath(S_2-S_2^*))$ with $\mu,\delta\not =0$. For this model the Chern number is equal to $\pm 1$ pending on the signs of $\mu$ and $\delta$, as shown {\it e.g.} in \cite{DDS}. This remains valid for small random $V$ and $W$. Furthermore, the Chern number is well-known to be equal to $\mbox{\rm Ind}(PFP)$.  Hence inserting a half-flux as in \eqref{eq-BdGentries} produces a zero energy mode. This is a discrete analog of \cite{RG}.
\hfill $\diamond$

\vspace{.2cm}

\noindent {\bf Example 2} The Wilson-Dirac operator from \cite{EF} can be written in the form
$$
H
\,=\,
\imath\,
\begin{pmatrix}
S_2-S_2^* & S_1-S_1^*+\mu +\lambda(4+S_1+S_1^*+S_2+S_2^*)
\\
S_1-S_1^*-\mu -\lambda(4+S_1+S_1^*+S_2+S_2^*) & -S_2+S_2^*
\end{pmatrix}
\,,
$$
with $\lambda,\mu\in\RM$. For $\frac{\mu}{\lambda}\neq0,4,8$ the system is gapped. This Hamiltonian is in the Majorana representation \eqref{eq-MajRep} of an operator with even PHS. For $\frac{\mu}{\lambda}\in(0,4)$  the Chern number is equal to  the sign of $-\lambda$ and for $\frac{\mu}{\lambda}\in(4,8)$ is equal to the sign of $\lambda$. For all other parameters the Chern number vanishes \cite{EF}. Two simple techniques to check these results are explained in \cite{DDS}. Now one can again insert a half-flux which then carries a zero mode. This provides a rigorous proof of \cite{EF}, but also shows the stability of the zero modes under perturbations ({\it e.g.} a small random potentials).
\hfill $\diamond$

\vspace{.2cm}

\noindent {\bf Example 3} As another lattice BdG model let us consider a $d+id$ wave superconductor (just one spin component of it). Again $h=S_1+S_1^*+S_2+S_2^*-\mu$ is the diagonal part of the BdG Hamiltonian, and $\Delta=\imath\delta(S_1+S_1^* -S_2-S_2^*+ (S_1-S_1^*)(S_2-S_2^*))$. Now $\Delta^t=\Delta$ so this model has an odd PHS (Class C). The Chern number and hence also $\mbox{\rm Ind}(PFP)$ is even, as shows a direct computation \cite{DDS}. Hence by Theorem~\ref{theo-PHS} there is no stable zero mode for $H_{\frac{1}{2}}$.
\hfill $\diamond$

\vspace{.2cm}

Actually the last example is a manifestation of the general fact that any  model from Class C has an even index $\Ind(PFP)$. Consequently there is no stable zero mode attached to a half-flux inserted in a Class C model, other than claimed in the literature \cite{Roy}.

\begin{theo} 
\label{theo-ClassC}
Let $H$ be a gapped two-dimensional {\rm BdG} Hamiltonian with odd {\rm PHS} with Fermi projection $P=\chi(H\leq 0)$. Then
$$
\mbox{\rm Ind}(PFP)\;\in\;2\,\ZM
\;.
$$
\end{theo}

\noindent {\bf Proof.} 
First of all, let us show that
$$
F\,:\,
\Ker(PFP)
\,\to\,
\Ker((\one-P)F^*(\one-P))
\;,
$$
is a unitary map. Indeed, let $\psi_1,\ldots,\psi_N$ be an orthonormal basis of $\Ker(PFP)$.
Then $\psi_n=P\psi_n$ implies $(\one-P)F\psi_n=F\psi_n$ so that $F\psi_n\in(\one-P)\Hh$. Hence
$$
(\one-P)F^*(\one-P)F\psi_n\;=\;
(\one-P)F^*F\psi_n
\;=\;(\one-P)\psi_n
\;=\;0
\;.
$$
Hence $F\psi_1,\ldots,F\psi_N$ is an orthonormal set in $\Ker((\one-P)F^*(\one-P))$. This argument can be reversed showing that $F$ is a bijection.

\vspace{.1cm}

Second of all, the hypothesis on $H$ implies that for any odd function $f:\RM\to\RM$ one has $\KPH^*\overline{f(H)}\KPH=-f(H)$. Therefore the Fermi projection satisfies $\KPH^*\overline{P}\KPH=\one -P$. Consequently, using $\overline{F}=F^*$ and $[\KPH,F]=0$,
$$
\Ker(PFP)
\;=\;
\KPH\;\Ker(\KPH PFP\KPH)
\;=\;
\KPH\;\overline{\Ker\bigl((\one-P)F^*(\one-P) \bigr)}
\;.
$$
Hence if $\Cc$ denotes the complex conjugation, then this establishes an anti-linear bijection
$$
\KPH\,\Cc\,:\,\Ker(PFP)
\,\to\,
\Ker((\one-P)F^*(\one-P))
\;.
$$
Combining the two maps, one obtains an anti-unitary $F^*\KPH\,\Cc\,:\,\Ker(PFP)\to\Ker(PFP)$ which satisfies $(F^*\KPH\,\Cc)^2=-\one$, namely $F^*\KPH\,\Cc$ is a quaternionic structure on the finite dimensional vector space $\Ker(PFP)$. It follows that $\Ker(PFP)$ is even dimensional (an explicit argument for this conclusion uses that $F^*\KPH$ is a skew-symmetric unitary matrix on $\Ker(PFP)$). The same argument also implies that $\Ker(PF^*P)$ is even dimensional and therefore also $\Ind(PFP)$ is even. Let us point out that for even PHS all arguments transpose, except that $(F^*\KPH\,\Cc)^2=\one$ so that there is no restriction on the dimension of $\Ker(PFP)$.
\hfill $\Box$


\subsection{Time reversal symmetric systems}
\label{sec-TRS}

The class of Hamiltonians on $\Hh=\ell^2(\ZM^2)\otimes\CM^L$ with $\CM^L=\CM^N\otimes\CM^{2s+1}$ considered here are of the form
\begin{equation}
\label{eq-disorderedHamgen}
H_\alpha
\;=\;
\sum_{|n|\leq R}\,  t_{n}\;(S_1^{B,\alpha})^{n_1}(S_2^{B,\alpha})^{n_2}
\;+\;
V
\;,
\end{equation}
where the $t_n$ are now compact operators with values in the $L\times L$ matrices satisfying \eqref{eq-hoppingcond}, and the potential $V=\sum_{n\in\ZM^2}v_n\,|n\rangle\langle n|$ also has coefficients $v_n$ in the self-adjoint $L\times L$ matrices. For $\alpha=0$, the notation $H=H_0$ will be used as above. Hence $H_\alpha$ is nothing, but a matrix-valued version of \eqref{eq-disorderedHarper}. This framework allows to model operators on the hexagon lattice with spin orbit coupling terms, beneath others \cite{ASV}. In particular, a disordered Kane-Mele model \cite{KM} is in this class (with $N=2$ and $s=\frac{1}{2}$). 

\vspace{.2cm}

\begin{figure}
\begin{center}
\includegraphics[height=6cm]{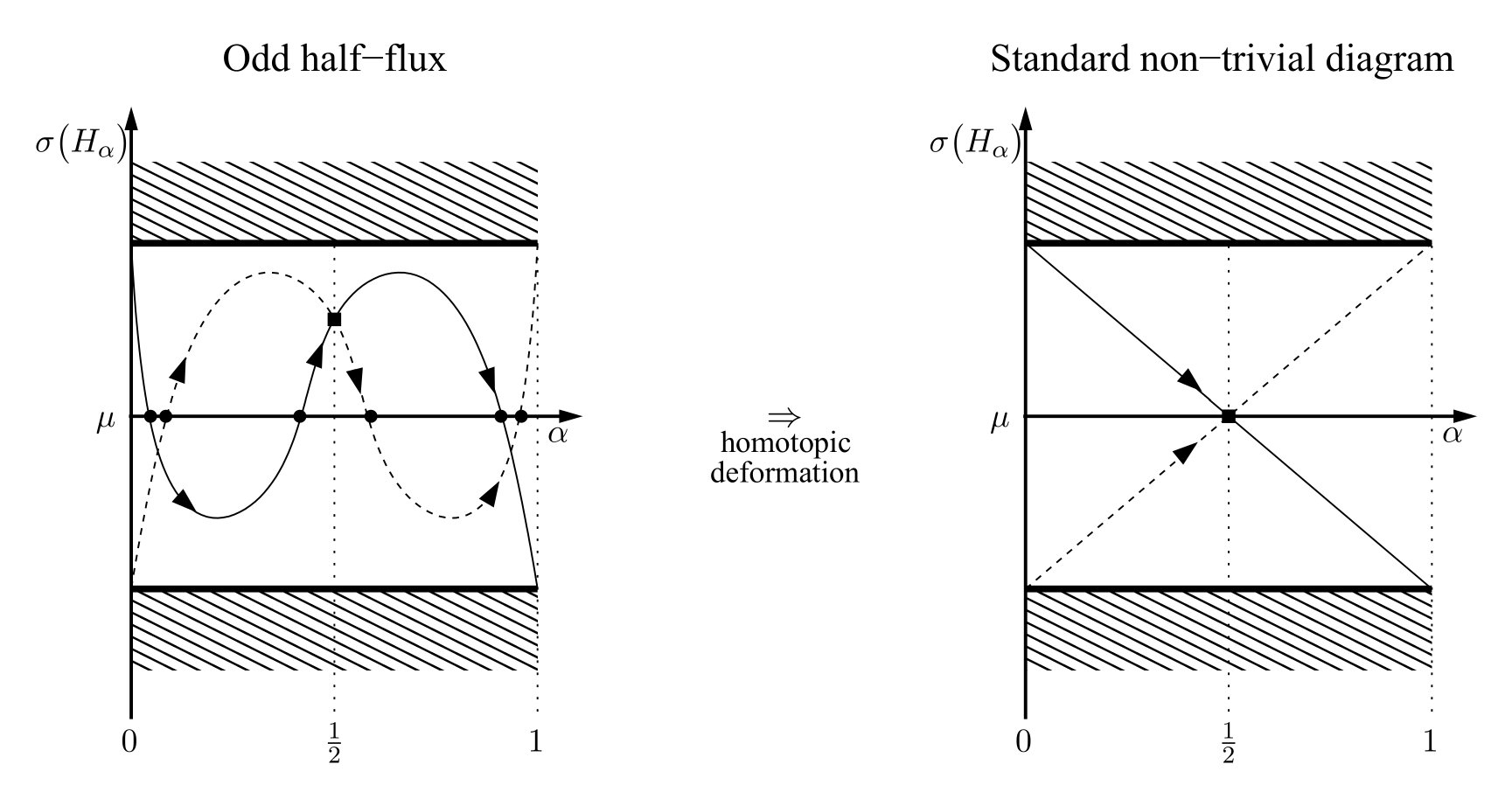}
\caption{\it 
Typical example of spectral flows associated with flux insertions $\alpha\mapsto H_\alpha$ subjected to an odd {\rm TRS} with non-trivial invariant $\Ind_2(PFP)=1$. By homotopy within the spectral flows with odd {\rm TRS}  it can be deformed to a cross with spectral multiplicity $1$, except for the Kramers doublet at $\alpha=\frac{1}{2}$. Note that the Kramers doublet need not lie at the Fermi level as the right figure may suggest.
\label{fig-TRS1}}
\end{center}
\end{figure}

\begin{figure}
\begin{center}
\includegraphics[height=6cm]{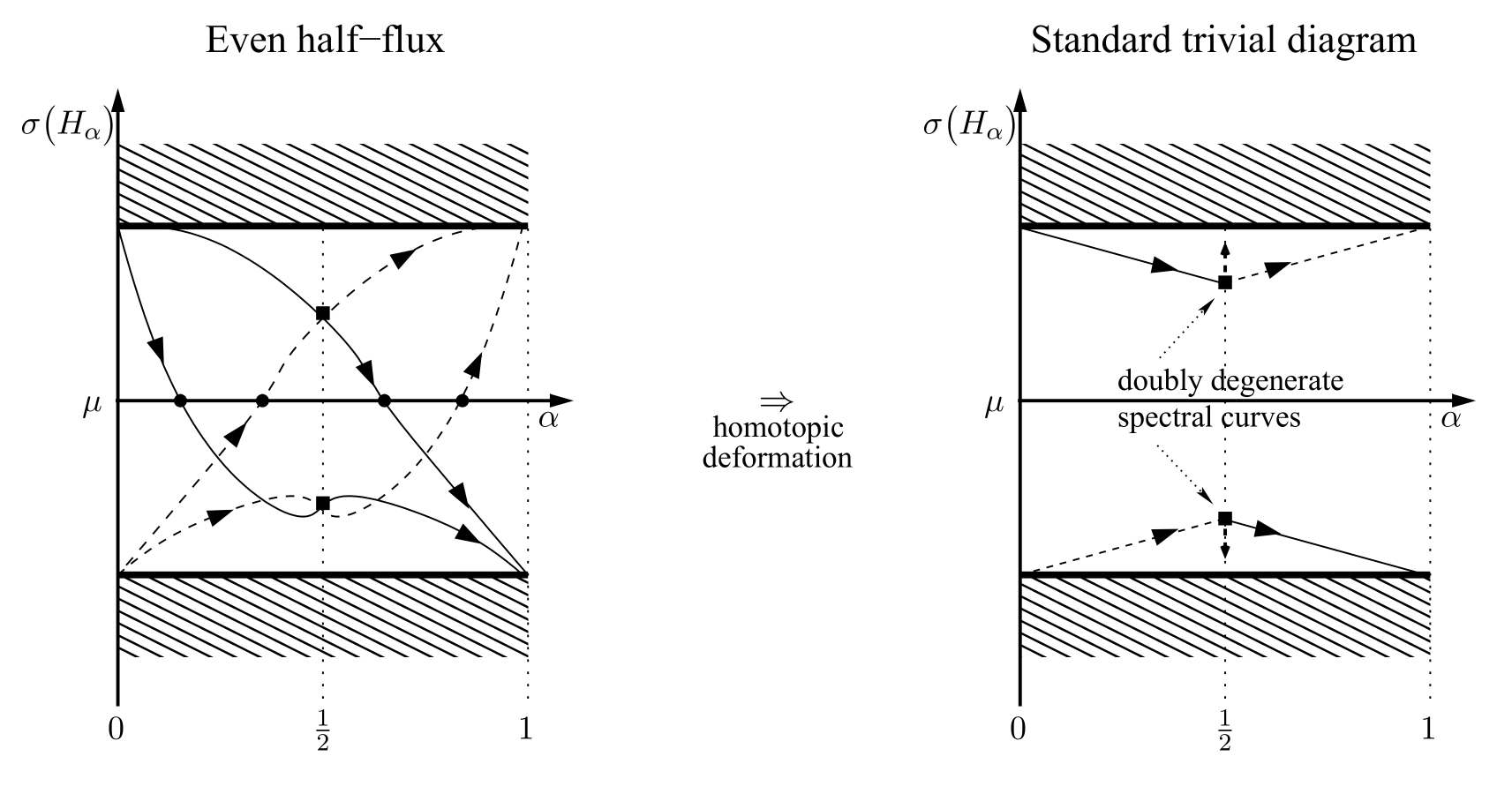}
\caption{\it 
Typical example of spectral flows associated with flux insertions $\alpha\mapsto H_\alpha$ subjected to an odd {\rm TRS} with non-trivial invariant $\Ind_2(PFP)=1$. By homotopy within the spectral flows with odd {\rm TRS} it can be deformed into a trivial spectral flow not having any spectrum an the Fermi level. 
\label{fig-TRS0}}
\end{center}
\end{figure}

By \eqref{eq-magtranscomplexconj} one has $\overline{S^{B,\alpha}}=S^{-B,-\alpha}$, so that TRS of $H$ imposes $B=0$. Let now $P=\chi(H\leq \mu)$ be the Fermi projection associated to a Fermi energy $\mu\in\RM$ lying in a gap of $H$. It then satisfies $I_{\mbox{\rm\tiny tr}}^*P^tI_{\mbox{\rm\tiny tr}}=P$ where $A^t=(\overline{A})^*$ denotes the transpose of an operator. This implies that also the Fredholm operator $PFP$ on $P\Hh$ satisfies 
$$
I_{\mbox{\rm\tiny tr}}^*\,(PFP)^t\,I_{\mbox{\rm\tiny tr}}
\;=\;
PFP
\;.
$$ 
This implies that kernel and cokernel of the Fredholm operator $T=PFP$ are of same dimension and consequently the index of $PFP$ vanishes. For integer $s$, one has $(I_{\mbox{\rm\tiny tr}})^2=\one$ so that the operator $T$ lies in the set of so-called  even symmetric Fredholm operators which is path-connected \cite{Sch}. Therefore it will be supposed from now on that $s$ is half-integer and that $H$ has odd TRS \eqref{eq-TRS}, in particular, $B=0$. In this case $(I_{\mbox{\rm\tiny tr}})^2=-\one$ and $T$ is an odd-symmetric Fredholm operator and has a well-defined homotopy invariant \cite{Sch} given by
$$
\Ind_2(PFP)
\;=\;
\dim(\ker(PFP))\,\mbox{\rm mod} \;2
\;\in\;\ZM_2
\;.
$$
In the following, one consequence of a non-trivial $\ZM_2$ index is discussed, namely the existence of a Kramers' degenerate bound state for a half-flux Hamiltonian associated to $H$. As $I_{\mbox{\rm\tiny tr}}^*\,\overline{S^{0,\alpha}_j}\,I_{\mbox{\rm\tiny tr}}={S^{{0,-\alpha}}_j}$ and $H$ has TRS, it follows that 
\begin{equation}
\label{eq-TRSsym}
I_{\mbox{\rm\tiny tr}}^*\,\overline{H_\alpha}\,I_{\mbox{\rm\tiny tr}}
\;=\;
H_{-\alpha}
\;=\;
F^*\,H_{1-\alpha}\,F
\;.
\end{equation}
This shows that $H_\alpha$ for $\alpha\not =0$ does not have TRS in the sense of \eqref{eq-TRS}, but at $\alpha=\frac{1}{2}$ one has 
$$
(I_{\mbox{\rm\tiny tr}}F^*)^*\,\overline{H_{\frac{1}{2}}}\,(I_{\mbox{\rm\tiny tr}}F^*)
\;=\;
H_{\frac{1}{2}}
\;.
$$
This resembles an odd TRS, but the unitary $V=I_{\mbox{\rm\tiny tr}}F^*$ is neither real nor does it square to a multiple of the identity. On the other hand, it satisfies $\overline{V}\,V=-\one$ and this is sufficient to run the Kramers' degeneracy argument in the proof of Proposition~\ref{prop-Kramersdeg}. Of course, the Kramers' degeneracy also holds for $H_0$ (and $H_1$). Furthermore, one can immediately read off from \eqref{eq-TRSsym} that the spectral curves of $\alpha\in[0,1]\mapsto H_\alpha$ have a reflection property about $\alpha=\frac{1}{2}$. Combined these facts imply that there are two classes of spectral flows  which cannot be deformed into each other \cite{ASV,DS0} within the set of spectral curves having the symmetry \eqref{eq-TRSsym} and Kramers' degeneracy at $\alpha=0,\frac{1}{2},1$. Figures~\ref{fig-TRS1} and \ref{fig-TRS0} give examples for each of the two classes. The invariant distinguishing between them is denoted by $\SF_2(\alpha\in[0,1]\mapsto H_\alpha)\in\ZM_2$. One way to define this $\ZM_2$ invariant is as the number of spectral intersection of $\alpha\in[0,\frac{1}{2}]\mapsto H_\alpha$ through some (any) $\mu$ in the gap modulo $2$.  The following result shows that the $\ZM_2$ index $\Ind_2(PFP)$ allows to predict which spectral flow one has.

\begin{theo} 
\label{theo-Z2TRS} 
Suppose that $\mu\in\RM$ lies in a gap of the Hamiltonian $H$ which has odd {\rm TRS}. Then
$$
\SF_2(\alpha\in[0,1]\mapsto H_\alpha)
\;=\;
\Ind_2(PFP)
\;.
$$
Hence, if $\Ind_2(PFP)=1$, the Hamiltonian $H_{\frac{1}{2}}$ has an odd number of Kramers' degenerate eigenvalues in the gap and, in particular, at least one of them.
\end{theo}

\noindent {\bf Proof.}
As in the proof of Theorem~\ref{theo-LaughlinFlux}, let $g:\RM\to [0,1]$ be a non-increasing function such that $P=g(H_0)$ and set $P_\alpha=g(H_\alpha)$.  Again by Proposition~\ref{prop-fluxprepare}(v) the operators $P_\alpha-P$ are compact and furthermore \eqref{eq-TRSsym} implies $P_{1-\alpha}=(I_{\mbox{\rm\tiny tr}}F^*)^*(P_\alpha)^tI_{\mbox{\rm\tiny tr}}F^*$. Hence all the hypothesis of Theorem~\ref{theo-Z2Flow} are satisfied (with $T=PFP$ and $U_T=F$). Thus the result follows.
\hfill $\Box$

\vspace{.2cm}

\noindent {\bf Example} It is well-known that the Kane-Mele model is an example with a non-trivial $\ZM_2$ index and that this model can be obtained by perturbing a direct sum of two Haldane models \cite{KM}. Let us slightly generalize this construction and calculate the associated $\ZM_2$ index. Suppose given a two-dimensional tight-binding Hamiltonian $h$ on $\ell^2(\ZM^2)\otimes \CM^N$ for which $\Ind(pFp)$ with $p=\chi(h\leq \mu)$ is odd, but for which the constant magnetic field $B$ vanishes (for example, a Haldane model). Then let us consider a Hamiltonian on $\Hh=\ell^2(\ZM^2)\otimes \CM^N\otimes \CM^2$ (that is, $s=\frac{1}{2}$ and $I_{\mbox{\rm\tiny tr}}=\binom{0 \;-1}{1\;\;\;0}$) which is of the form
\begin{equation}
\label{eq-KMdouble}
H
\;=\;
\begin{pmatrix} h & g \\ g^* & \overline{h}
\end{pmatrix}
\;,
\end{equation}
in the grading of the spin degree of freedom $\CM^2$. This operator is of the form \eqref{eq-disorderedHamgen} with $B=0$. The TRS $(I_{\mbox{\rm\tiny tr}})^*\overline{H}I_{\mbox{\rm\tiny tr}}=H$ is guaranteed by $g^t=-g$. In the case of the Kane-Mele model, the operator $g$ is essentially the Rashba coupling. If $g$ vanishes, the system decouples into a direct sum (one component of the spin is conserved) and in this situation the $\ZM_2$ index will now be calculated. By homotopy invariance of the $\ZM_2$ index, it then remains constant as $g$ is homotopically added, as long as the gap remains open. For $g=0$, the Fermi projection $P=\chi(H\leq \mu)$ is given by $P=\binom{p\;\;0}{0\;\;\overline{p}}$. Therefore
$$
PFP
\;=\;
\begin{pmatrix}
p\,F\,p & 0 \\
0 & \overline{p}\,F\,\overline{p}
\end{pmatrix}
\;.
$$
Now $\Ind(pFp)$ is odd by hypothesis. Therefore, if $\dim(\Ker(pFp))$ is odd (resp. even), then $\dim(\Ker(pF^*p))=\dim(\Ker(p\,\overline{F}\,p))=\dim(\Ker(\overline{p}\,F\,\overline{p}))$ is even (resp. odd). It follows that $\dim(\Ker(PFP))$ is indeed odd so that $\Ind_2(PFP)=1$. If $\Ind(pFp)$ is even, then $\Ind_2(PFP)=0$ by the same argument. Let us point out that in the context of periodic operators this conclusion agrees with  Corollary 4.12 of \cite{DG} which relates the value of the  $\ZM_2$ index of $P$ with the parity of the Chern number $\Ind(pFp)$ associated with the reduced projection $p$.
\hfill $\diamond$

\subsection{Hamiltonians with chiral symmetry}
\label{sec-chiral}

Now the Hamiltonian satisfies \eqref{eq-SLS} and the Fermi level $\mu=0$ lies in a gap.  This implies $\KSL^*P\KSL=\one-P$ for the Fermi projection so that $\Ind(PFP)=\Ind((\one-P)F(\one-P))$ as long as $[\KSL,F]=0$ (which is supposed from now on). On the other hand, the identity 
$$
F
\;=\;
\begin{pmatrix}
PFP & PF(\one-P) \\
(\one-P)FP & (\one-P)F(\one-P)
\end{pmatrix}
\;,
$$
combined with the fact that $PF(\one-P)=P[F,P]$ is compact implies $\Ind(PFP)+\Ind((\one-P)F(\one-P))=0$. Thus one concludes $\Ind(PFP)=0$. By Theorem~\ref{theo-LaughlinFlux} (more precisely the matrix valued version of it) there is no spectral flow associated to a family $\alpha\in[0,1]\mapsto H_\alpha$ from  $H_0=H$ to $H_1=FHF^*$ such that $H_\alpha-H$ is compact. Such a family is again realized by inserting a flux tube. This agrees with the vanishing entry in the line AIII of Table~\ref{table1}.

\vspace{.2cm}

In the CAZ classification of Table~\ref{table1} there are four other classes which have a chiral symmetry combined with other symmetries, namely classes DIII, CII, CI, BDI.  In Classes CI and BDI, the even TRS does not lead to secondary invariants, as explained in Section~\ref{sec-TRS}. However, the odd TRS in Classes DIII and BDI allows in principle for non-trivial secondary $\ZM_2$ invariants. Actually in DIII such a non-trivial model can be constructed as shown next.

\vspace{.2cm}

\noindent {\bf Example} 
The basic idea to construct a non-trivial model in Class DIII is similar to the Kane-Mele model which is a perturbation of two copies of the Haldane model. Hence let $h$ be a $p+ip$ wave BdG model on $\ell^2(\ZM^2)\otimes \CM^2_\ph$ with vanishing constant magnetic field. Its Fermi projection $p=\chi(h\leq 0)$ satisfies $\Ind(pFp)=1$ for adequate values of the parameters, see Section~\ref{sec-PHS}. Then $\overline{h}$ is a $p-ip$ BdG model with $\Ind(\overline{p}\,F\,\overline{p})=-1$. Now $H$ acting on  $\ell^2(\ZM^2)\otimes \CM^2_\ph\otimes \CM^2_{\mbox{\rm\tiny tr}}$ is given by \eqref{eq-KMdouble}. To insure odd TRS, it will again be supposed that $g^t=-g$. Furthermore, even PHS is given if $\KPH^*\overline{g} \KPH =-g$ which is also supposed from now on. Now by the same argument as in Section~\ref{sec-TRS} one concludes $\Ind_2(PFP)=1$ for $P=\chi(H\leq 0)$. This non-trivial value is conserved for small $g$.
\hfill $\diamond$

\vspace{.2cm}

In principle, this construction can be carried out in exactly the same way for Class CII, but as the building blocks $h$ are  necessarily in Class C, their indices are always even by Theorem~\ref{theo-ClassC} so that no models with non-trivial $\ZM_2$ indices can be obtained. This gives the corresponding entry in Table~\ref{table1}. Combining the arguments of Sections~\ref{sec-PHS} and \ref{sec-TRS}, one now obtains the following result.

\begin{theo} 
\label{theo-DIII} 
Let $H=H_0$ be a two-dimensional {\rm BdG} Hamiltonian with even {\rm PHS} and odd {\rm TRS}, namely $H$ is in Class {\rm DIII}. Suppose that $0$ lies in a gap of $H$  and let $P=\chi(H\leq 0)$ be the Fermi projection. Let $H_{\frac{1}{2}}$ the Hamiltonian obtained by setting $\alpha=\frac{1}{2}$ in a family of the form {\rm \eqref{eq-BdGalpha}} and {\rm \eqref{eq-BdGalpha2}}. If $\mbox{\rm Ind}_2(PFP)=1$, then $H_{\frac{1}{2}}$ has a Kramers degenerate zero mode that is stable under perturbations of $H$.
\end{theo}

\appendix

\section{Spectral flow of unitary dilations}
\label{sec-SFdilations}

This appendix presents the main results of Phillips \cite{Phi} and the companion paper \cite{DS0} in a form adapted to the applications in the main text. An intuitive definition of spectral flow is given in Section~\ref{sec-Laughlin}. For a more detailed (and general) definition, the reader is referred to \cite{Phi,DS0}. 


\begin{theo}
\label{theo-FredFlow} {\rm \cite{Phi}} Let $T\in \Bb(\Kk)$ with $\|T\|\leq 1$ be a bounded Fredholm operator on a Hilbert space $\Kk$ and let $U_T\in\Bb(\Hh)$ be an arbitrary unitary dilation, namely there is injective partial isometry $\Pi:\Kk\hookrightarrow\Hh$ into another Hilbert space $\Hh$ such that $T=\Pi^* U_T\Pi$. Associated is a projection $P=\Pi\Pi^*$ on $\Hh$. Let $\alpha\in[0,1]\mapsto P_\alpha$ be any continuous path of self-adjoint operators from $P_0=P$ to $P_1=U_T^*PU_T$ such that $P_\alpha-P$ is a compact operator. Then the spectral flow associated to this path through any spectral point $\mu\in(0,1)$ is linked to the index via
\begin{equation}
\label{eq-SFInd}
\Ind(T)\;=\;-\;\SF\bigl(\alpha\in[0,1]\mapsto P_\alpha\;\;\mbox{\rm by }\mu \bigr)
\;.
\end{equation}
\end{theo}

Two different proofs of Theorem~\ref{theo-FredFlow} are also contained in \cite{DS0}. Next let us recall some definitions and facts from \cite{Sch}. Let $I$ be a real unitary on a Hilbert space $\Kk$ satisfying $I^2=-\one$. A bounded operator $T\in\Bb(\Kk)$ is called odd symmetric if and only if $I^*T^tI=T$. The set of odd symmetric Fredholm operators contains two connected components labelled by the homotopy invariant $\Ind_2(T)=\dim(\Ker(T))\!\!\mod 2\in\ZM_2$. Moreover, for an odd symmetric compact operator $K$, one has $\Ind_2(T+K)=\Ind_2(T)$. The following result, based on a spectral flow argument, follows by combining Theorem~7 of \cite{DS0}.

\begin{theo}
\label{theo-Z2Flow} {\rm \cite{DS0}} Let $T\in \Bb(\Kk)$ with $\|T\|\leq 1$ be a bounded odd symmetric Fredholm operator on a Hilbert space $\Kk$. Let $U_T\in\Bb(\Hh)$ be an arbitrary odd symmetric unitary dilation, namely a unitary dilation as above for which $I$ extends to $\Hh$ and $U_T$ is odd symmetric. Set again $P=\Pi\Pi^*$. Let $\alpha\in[0,1]\mapsto P_\alpha$ be any continuous path of self-adjoint operators from $P_0=P$ to $P_1=U_T^*PU_T$ such that $P_\alpha-P$ is a compact operator and 
$$
P_{1-\alpha}\;=\;
(IU_T)^*(P_\alpha)^t(IU_T)
\;.
$$
Then the $\ZM_2$-valued spectral flow by any $\mu\in(0,1)$, defined as in {\rm Section~\ref{sec-TRS}} and {\rm \cite{DS0}}, satisfies
$$
\SF_2(\alpha\in[0,1]\mapsto P_\alpha\;\;\mbox{\rm by }\mu )
\;=\;
\Ind_2(T)
\;.
$$
\end{theo}

\section{Toeplitz extension of the rotation algebra}
\label{sec-rotation}

Let $\Aa_B=C^\ast(S^B_1,S^B_2)$ be the C$^\ast$-algebra generated by the (Zak) magnetic translations with constant  magnetic field $B\in\RM$ as defined in Section~\ref{sec-magtrans}. Hence $\Aa_B$ is a subalgebra of the bounded operators on $\ell^2(\ZM^2)$. We will refer to $\Aa_B$ simply as the rotation algebra because it is known to be a faithful representation of the more abstract version of it. An important property of $\Aa_B$ is that it has a trivial intersection with the ideal $\Kk$ of compact operators on $\ell^2(\ZM^2)$, namely  $\Aa_B\cap\Kk=\{0\}$. This can be proved observing that $\Aa_B$ is invariant under a (projective) $\ZM^2$ action implemented by a pair of independent  dual magnetic translations. This implies that the eigenspaces of elements  in $\Aa_B$ are necessarily infinitely degenerate and so compact operators cannot be in $\Aa_B$. Furthermore let $\Tt(\Aa_B)=C^\ast(S^{B}_1,S^{B}_2,P_0)$ be the C$^\ast$-algebra generated by the magnetic translations and the one-dimensional projection $P_0=|0\rangle\langle 0|$. Evidently $\Aa_B\subset\Tt(\Aa_B)$. However, also $\Kk\subset\Tt(\Aa_B)$. Indeed all the rank $1$ operators $|n\rangle\langle m|$ can be generated in $\Tt(\Aa_B)$ with the application of the shifts  $S^{B}_1$ and $S^{B}_2$ to $P_0$ and the rank 1 operators are norm-dense in $\Kk$. Thus there are canonical (injective) inclusions by
$$
\imath\;:\;\Kk\; \hookrightarrow\;\Tt(\Aa_B)\;,\qquad
\qquad
\jmath\;:\;\Aa_B\; \hookrightarrow\;\Tt(\Aa_B)\;.
$$

 \begin{proposi}
 \label{cor:uniq}
Each $A\in \Tt(\Aa_B)$ decomposes uniquely as $A=A_\infty+A_{0}$ with $A_\infty\in\Aa_{B}$ and $A_{0}\in \Kk$. 
\end{proposi}

\noindent {\bf Proof.} 
The existence of the decomposition  follows from the fact that $\Tt(\Aa_B)$ is generated by the closed sub-algebras $\Aa_{B}$ and $\Kk$ and  $\Kk$ is a two-sided ideal. For the unicity, let us assume that  $A=A_\infty+A_{0}=A_\infty'+A_{0}'$. then $A_\infty-A'_\infty=A_{0}'-A_{0}$. Since $\Aa_{B}$ has no non-trivial compacts it follows that $A_\infty-A'_\infty=0$, so that also $A_{0}'-A_{0}=0$. 
\hfill $\Box$

\vspace{.2cm}

This decomposition property for elements in $\Tt(\Aa_B)$ allows us to define the surjective C$^\ast$-homomorphism
$$
A\in \Tt(\Aa_B)\;\overset{{\rm ev}}{\longrightarrow}\; A_\infty\;\in\;\Aa_{B}\;.
$$
%

\begin{theo}
The sequence 
\begin{equation}
\label{eq:toep_ext1}
0\;\longrightarrow\;\Kk\;\stackrel{\imath}{\hookrightarrow}\;\Tt(\Aa_B)\; 
\mathop{\overset{{\rm ev}}{\rightarrow}}_{\underset{{\jmath}}{\hookleftarrow}}
%
\;\Aa_B\;\longrightarrow\;0\;
\end{equation}
is exact and right split. 
\end{theo}
 
 \noindent {\bf Proof.} 
The exactness is a consequence of the injectivity of $\imath$ and the surjectivity of ${\rm ev}$. The splitting property follows observing that ${\rm ev}\circ \jmath={\rm Id}_{\Aa_{B}}$. 
\hfill $\Box$

\vspace{.2cm}

According to the standard terminology, $\Tt(\Aa_B)$ is said to be a \emph{trivial extension} of $\Aa_{B}$ by the compacts and the map $\jmath$ is called the \emph{lifting map}. Since $\Kk$ is a closed two-sided ideal in $\Tt(\Aa_B)$,  the quotient $\Tt(\Aa_B)/ \Kk$ is a C$^\ast$-algebra and  one has a C$^\ast$-algebra isomorphism $ \Aa_B\,{\simeq}\, \Tt(\Aa_B)/ \Kk$. This isomorphism implies
\begin{equation}
\label{eq:ext_02}
\|A_\infty\|\;=\;\inf_{K\in\Kk}\|A_\infty+K\|
\;,
\qquad\qquad\forall\ \ A_\infty\in\Aa_B\;,
\end{equation}
namely compact perturbations cannot decrease the norm of any element of the algebra $\Aa_B$. 

\vspace{.2cm}

Recall that the Toeplitz algebra of Toeplitz operators acting on Hardy space is generated by a shift and a one-dimensional projection (all on $\ell^2(\ZM)$).  By analogy, we call $\Tt(\Aa_B)$ a Toeplitz extension of $\Aa_B$. Let us point out that a different type of Toeplitz extension of the rotation algebra was studied in \cite{Par}.

\vspace{.2cm}

\noindent {\bf Remark} 
Equation \eqref{eq:ext_02} implies $\|{\rm ev}(A)\|\leqslant \|A\|$ for all  $A\in \Tt(\Aa_B)$, {\it i.e.} the continuity of the $C^\ast$-morphism ${\rm ev}$. This is the best that one can do since for a compact $K\neq0$, one has $\|K\|>0$ and $\|{\rm ev}(K)\|=0$. This fact differs from the usual Toeplitz extension where the norm of a Toeplitz operator is equal to the norm of its symbol, hence ${\rm ev}$ is an isometry in this case. Also the extension of $\Aa_B$ considered in \cite{Par} preserves the norm (by Proposition~1 in \cite{Par}).
\hfill $\diamond$
 
\vspace{.2cm}
 
The following result implies that the Hamiltonians $H_\alpha(0)$ defined in \eqref{eq-disorderedHarper}, but with $\lambda=0$, lie in the Toeplitz extension.

\begin{theo}
\label{theo-pairings2} 
Let  $C^\ast(S^{B,\alpha}_1,S^{B,\alpha}_2)$ be the C$^\ast$-algebra generated by $S^{B,\alpha}_1$ and $S^{B,\alpha}_2$, namely the two magnetic translations twisted by the insertion of the  Aharonov-Bohm flux $\alpha\in\RM$. Then:

\vspace{.1cm}

\noindent {\rm (i)} For all $n\in\ZM$, the $C^\ast$-algebra $C^\ast(S^{B,n}_1,S^{B,n}_2)$ is unitarily equivalent to $\Aa_B$. In particular,  

$C^\ast(S^{B,n}_1,S^{B,n}_2)\cap\Kk=\{0\}$.

\vspace{.1cm}

\noindent {\rm (ii)}  $\Tt(\Aa_B)=C^\ast(S^{B,\alpha}_1,S^{B,\alpha}_2)$ for all $\alpha\in\RM\setminus\ZM$.
\end{theo}

\noindent {\bf Proof.}  
(i) follows observing that $S^{B,0}_j=S^{B}_j$ and $S^{B,\alpha+1}_j=FS^{B,\alpha}_jF^\ast$ where $F$ is the unitary defined in Proposition \ref{prop-magtransprop}. (ii) Let us start by proving that $S^{B}_1,S^{B}_2,P_0\in C^\ast(S^{B,\alpha}_1,S^{B,\alpha}_2)$. First of all, a direct computation shows that $e^{-\imath B}\;S^{B,\alpha}_1S^{B,\alpha}_2(S^{B,\alpha}_1)^\ast(S^{B,\alpha}_2)^\ast-\one\propto P_0$. Second of all, Proposition~\ref{prop-magtransprop} assures $S^{B}_j=S^{B,\alpha}_j-K_j^{B,\alpha}$ with $K_j^{B,\alpha}\in\Kk$ and thus $S^{B}_1$ and $S^{B}_2$ are also in  $C^\ast(S^{B,\alpha}_1,S^{B,\alpha}_2)$. Hence $\Tt(\Aa_B)\subset C^\ast(S^{B,\alpha}_1,S^{B,\alpha}_2)$. The opposite inclusion follows from $\Kk\subset \Tt(\Aa_B)$, so that also $S^{B,\alpha}_j=S^{B}_j+K_j^{B,\alpha}\in\Tt(\Aa_B)$.
\hfill $\Box$

\vspace{.2cm}

Finally let us calculate the $K$-theory of the Toeplitz extension. The exact sequence \eqref{eq:toep_ext1} produces the usual cyclic six term exact sequence
$$
\begin{aligned}\label{eq:exact_mat3}
\ZM=& K_0(\Kk) & \overset{\imath_\ast}{\longrightarrow} & & K_0(\Tt(\Aa_B)) & & \overset{{\rm ev}_\ast}{\longrightarrow}& & K_0(\Aa_B)&
\\
\\
& \Ind\;\;\uparrow  & & & & & & & \downarrow\;\;\exp&
\\
\\
& K_1(\Aa_B)  & \stackrel{\;\;{\mbox{\rm\tiny ev}_*}}{\longleftarrow} & & K_1(\Tt(\Aa_B)) & & \overset{\imath_\ast}{\longleftarrow} & & K_1(\Kk)&=0
\end{aligned}
$$
The fact that the connecting map $\exp$ is identically zero also follows from splitting property of \eqref{eq:toep_ext1} which implies that any projection $P\in\Aa_{B}$ (or matrix algebras over $\Aa_B$) lifts to a projection $\jmath(P)\in\Tt(\Aa_B)$ (or a matrix algebra over $\Tt(\Aa_B)$. For the same reason also the index map $\Ind$ is identically zero. Hence the six term sequence splits into two parts
\begin{equation}
\label{eq:exact_split-K1}
0\;  \longrightarrow\; K_j(\Kk) \;\overset{\imath_\ast}{\longrightarrow} \;K_j(\Tt(\Aa_B)) \;   \overset{{\rm ev}_\ast}{\longrightarrow} K_j(\Aa_{B})\; \longrightarrow\;0\;, \qquad j=0,1\;.
\end{equation}
In particular, one has $K_j(\Tt(\Aa_B)) \simeq K_j(\Kk) \oplus K_j(\Aa_{B})$. As $K_j(\Aa_B)$ is known \cite{PV}, this allows to compute the $K$-theory of $\Tt(\Aa_B)$.

\begin{theo}\label{theo:K-theory2}
Independently of $B\in\RM$, the $K$-theory of $\Tt(\Aa_B)$ is given by
$$
\begin{aligned}
&K_0(\Tt(\Aa_B))&\simeq\ \  &\ZM[P_0]\;\oplus\;\ZM[\one]\;\oplus\;\ZM[P_B]&\simeq\ \ &\ZM^3\\
&K_1(\Tt(\Aa_B))&\simeq\ \  &\ZM[S_1^B]\;\oplus\;\ZM[S_2^B]&\simeq\ \ &\ZM^2
\end{aligned}
$$
where $P_B\in \Aa_B$ is any \emph{Powers-Rieffel} projection. In particular, all the generators of the $K$-theory of  $\Tt(\Aa_B)$ can be chosen inside the C$^\ast$-algebra. 
\end{theo}

\noindent {\bf Proof.}  
Since $K_1(\Kk)=0$, it follows that $K_1(\Tt(\Aa_B))$ agrees with $K_1(\Aa_{B})$ which is generated by $[S_1^B]$ and $[S_2^B]$ as proved in  \cite[Corollary 2.5]{PV}. On the other hand, $K_0(\Tt(\Aa_B))\simeq K_0(\Kk)\oplus K_0(\Aa_{B})$. The first summand $K_0(\Kk)$ has generator $[P_0]$,  while the  two generators of $K_0(\Aa_{B})$ can  be chosen as $[\one]$ and $[P_B]$, see \cite[Appendix]{PV}. 
\hfill $\Box$

\vspace{.4cm}

\noindent {\bf Acknowledgements.} We thank Gian-Michele Graf for pointing us to the unpublished manu\-script of Macris \cite{Mac} and for making it available to us, and the referees for constructive comments that improved the manuscript. This work was partially funded by the DFG. G. De Nittis thanks the Humboldt Foundation for financial support.


\end{document}